\documentclass[11pt]{article}
\usepackage{amsfonts,latexsym,eucal,color,graphicx,epsfig,amssymb,amsmath,cite}

\newcommand{\be}{\begin{eqnarray}}
\newcommand{\ee}{\end{eqnarray}}
\newcommand{\pd}{\partial}
\newcommand{\RP}{Rayleigh-Plateau }
\newcommand{\GL}{Gregory-Laflamme }
\newcommand{\dd}{\mathrm{d}}
\newcommand{\UT}{\mathrm{UT}}
\newcommand{\NUT}{\mathrm{NUT}}
\newcommand{\SB}{\mathrm{SB}}
\newcommand{\rp}{\mathrm{RP}}

\definecolor{red  }{rgb}{1,0,0}
\definecolor{blue }{rgb}{0,0,1}
\definecolor{green}{rgb}{0,1,0}

\begin{document}

\begin{titlepage}

\begin{flushright}
{\small{arXiv:0811.2305 [hep-th]}}\\
{\small{WU-AP/294/08}}\\
\end{flushright}
\vspace{2cm}

\begin{center}
{\LARGE\sc Black Hole-Black String Phase Transitions \\[.5em]from Hydrodynamics}
\end{center}
\vspace{1cm}

\begin{center}
{\large Kei-ichi Maeda${}^\ast$ and Umpei Miyamoto${}^\dagger$}\\
\vspace{.4cm}
{\small{\textit{${}^\ast$Department of Physics, Waseda University,
Okubo 3-4-1, Tokyo 169-8555, Japan}}}\\
{\small{\textit{${}^\dagger$Racah Institute of Physics, Hebrew University, Givat Ram, Jerusalem 91904, Israel}}}\\
\vspace{0.4cm}
{\small{\tt{maeda@waseda.jp \quad umpei@phys.huji.ac.il}}}

\end{center}
\vspace{1.0cm}

\begin{abstract}
We discuss the phase transitions between three states of a plasma fluid (plasma ball, uniform plasma tube, and non-uniform plasma tube), which are dual to the corresponding finite energy black objects (black hole, uniform black string, and non-uniform black string) localized in an asymptotically locally AdS space. Adopting the equation of state for the fluid obtained by the Scherk-Schwarz compactification of a conformal field theory, we obtain axisymmetric static equilibrium states of the plasma fluid and draw the phase diagrams with their thermodynamical quantities. By use of the fluid/gravity correspondence, we predict the phase diagrams of the AdS black holes and strings on the gravity side. The thermodynamic phase diagrams of the AdS black holes and strings show many similarities to those of the black hole-black string system in a Kaluza-Klein vacuum. For instance, the critical dimension for the smooth transition from the uniform to non-uniform strings is the same as that in the Kaluza-Klein vacuum in the canonical ensemble. The analysis in this paper may provide a holographic understanding of the relation between the Rayleigh-Plateau and Gregory-Laflamme instabilities via the fluid/gravity correspondence.
\end{abstract}

\vspace{2.0cm}
\begin{flushleft}
\end{flushleft}

\end{titlepage}

\tableofcontents

\section{Introduction}
\label{sec:intro}

The AdS/CFT correspondence  provides one of the best ways to explore the 
nonperturbative regimes of a certain class of strongly coupled quantum field 
theories by working in classical gravity. 
Conversely, from the viewpoint of black hole physics, 
we may be able to understand strong gravitational phenomena 
by studying an appropriate limit of field theories. 
In this paper, we investigate
a class of AdS black holes,
in particular, their phase structures and possible phase transitions,
via the fluid/gravity correspondence
(see, e.g., \cite{fluid-gravity1,fluid-gravity2}), which has 
been proposed as a generalization of the AdS/CFT correspondence.

Recently, Aharony, Minwalla, and Wiseman~\cite{plasmaball} argued that
 in a class of large-$N$ gauge theories,
a plasma ball (i.e., a lump of `gluon' plasma) appears
via a first-order deconfinement phase transition 
above a critical temperature.
They also claimed that such a plasma ball will 
map to a finite energy black hole localized in the IR (infrared) region,
 and discussed what happens on the gravity side  
by studying various phenomena on the fluid side such 
as a plasma ball production and its subsequent decay by hadronization. 
In particular, they explicitly constructed a numerical domain wall solution to 
the Einstein equations with a negative cosmological constant which 
smoothly interpolates 
two spacetimes corresponding to the confined and deconfined 
phases. 
As a generalization of the analysis in~\cite{plasmaball}, Lahiri and 
Minwalla~\cite{plasmaring1} constructed a rotating plasma ring and a ball, 
which are dual to a rotating black ring and a rotating black hole, 
respectively (see also~\cite{plasmaring2}). By construction 
(which we will briefly review in the text),
 these gravity duals reside in the IR region of AdS space 
compactified on a Scherk-Schwarz circle. This kind of black ring and hole 
solutions have not been discovered yet. 
However, by comparing the phase diagrams 
of a plasma ring and plasma ball with those of a black ring and black hole in 
an asymptotically flat space,
they found qualitative agreements between them.

In this paper, we investigate the phase structures of another important class 
of black objects, i.e., black strings and black holes in a spacetime in
 which 
some spatial dimensions are compactified, from the  viewpoint of  
the fluid/gravity correspondence
(see~\cite{AMMW} for the analysis on a similar system from a 
holographic viewpoint). We solve the $d$-dimensional 
relativistic Navier-Stokes 
equations~\cite{fluid-gravity2} with an appropriate surface term to 
obtain static axisymmetric fluid equilibrium configurations.
The configurations contain a
spherical ball, a uniform tube, and a non-uniform tube.
To see what their gravity duals are, we recall the arguments in
\cite{plasmaball}. We consider the following solution to the $(d+2)$-dimensional
Einstein equations with a negative cosmological constant, 
$ R_{ab} = - (d+1)\ell^{-2} g_{ab} $:
\be
	ds_{d+2}^2
	=
	\ell^2
	\left(
		e^{2u}[ -dt^2 + T_{2\pi}(u)d\theta^2 + dw_i^2 ]
		+
		\frac{ 1 }{ T_{2\pi}(u) } du^2
	\right),
\label{eq:soliton}
\ee
where $i=1,2,\ldots,d-1$, and $\theta \in [0,2\pi)$. The function $T_x(u)$
is defined by 
\be
	T_x(u)
	=
	1 - \left[ \frac{x}{4\pi}(d+1)e^u \right]^{-(d+1)}.
\ee
This spacetime is regarded as an AdS${}_{d+2}$ 
space with a Scherk-Schwarz compactification, which
 is called the AdS soliton~\cite{Witten}. By taking $u\to \infty$, one recovers 
the AdS${}_{d+2}$  in the Poincar\'{e} coordinates with a uniform
circle of $\theta$. It should be noted that there is a cutoff in 
the IR region (i.e., small-$u$ region)
since the Scherk-Schwarz circle shrinks to a point 
at a finite value of $u$. 
Imposing a periodicity in the imaginary time, i.e.,  $\tau\equiv\tau+\beta$, 
($\tau=it$), 
one can regard this spacetime as a thermal gas of gravitons at 
temperature $T=\beta^{-1}$.
In addition, there exists another exact solution
\be
	ds^2_{d+2}
	=
	\ell^2
	\left(
		e^{2u}[ - T_{\beta}(u) dt^2 + d\theta^2 + dw_i^2 ]
		+
		\frac{ 1 }{ T_{\beta}(u) } du^2
	\right),
\label{eq:brane}
\ee
which has the same asymptotics as the spacetime (\ref{eq:soliton}).
This solution can be regarded 
as a black brane with temperature $T=\beta^{-1}$. Considering both spacetimes, 
Eqs.~(\ref{eq:soliton}) and (\ref{eq:brane}), together in an ensemble, one can 
show that in the low temperature regime ($T<T_c=1/2\pi)$ the AdS soliton 
(\ref{eq:soliton}) has a lower free energy, while in the high temperature 
regime ($T>T_c$) the black brane~(\ref{eq:brane}) has a lower free energy and 
dominates. Thus, the system undergoes a Hawking-Page type phase transition at 
$T=T_c$~\cite{HawkingPage,Witten}.

In~\cite{plasmaball,plasmaring1}, the plasma balls and plasma rings which extend 
uniformly in the $\theta$-direction and non-uniformly in the 
$w_i$-directions 
(which are assumed to be noncompact) were constructed on the UV (ultraviolet) boundary by 
solving the fluid equations\footnote{Note that the dual black holes and black rings 
also extend in the $u$-direction, although they are localized in this direction.}. In 
this paper, we compactify one of $w_i$'s on a circle, which we denote by 
coordinate $z$, and consider the plasma tube and plasma ball, where the former wraps
 the circle of the $z$-direction. As described in \cite{plasmaring1}, the 
horizon topology of a dual black object is obtained by fibering the plasma 
configuration with an $S^1$ that shrinks to zero size at the fluid 
edges\footnote{This condition on the shrinking of the $S^1$ circle corresponds to the 
fact that at the IR wall mapped from the outside of the plasma, the 
Scherk-Schwarz circle shrinks to zero size.}. Therefore, the plasma ball 
$B^{d-1}$ and plasma tube $B^{d-2}\times S^1$, which we will obtain in this
 paper, map to the black objects whose horizon topologies are $S^{d}$ and 
$S^{d-1} \times S^1$, respectively. Namely, they are a black hole and a black 
string, respectively, in an asymptotically 
AdS${}_d$ space on the Scherk-Schwarz circle 
$S_\theta$ and Kaluza-Klein circle $S_z$.

The gravity dual of the plasma tube obtained in this paper is rather 
different from the black string solutions in an asymptotically locally
AdS space which have been studied extensively in the
literature~\cite{AdS-strings}, 
as well as from the black strings in the asymptotically locally flat 
Kaluza-Klein space.
We will see, however, that the thermodynamic properties of our AdS black 
hole-black string system display various qualitative similarities to those 
of the black hole-black string system in the asymptotically locally flat 
Kaluza-Klein space. The phase transitions and their dimensional dependence 
will be elaborated in this paper.

This paper is organized as follows. In Sec.~\ref{sec:hydro}, we reduce the 
Navier-Stokes equations for a plasma fluid
to a hydrostatic equation for axisymmetric equilibrium 
states.
We then demonstrate that this equation is equivalent to that of 
Plateau's problem (or capillary minimizing problem), obtaining the constant mean curvature surfaces which model 
soap bubbles with a variational principle. In addition, we show that the uniform 
plasma tube is unstable against a perturbation, which is the fluid 
counterpart~\cite{Cardoso,MiyamotoMaeda,Miyamoto} of the Gregory-Laflamme instability~\cite{GL}.
In Sec.~\ref{sec:thermo}, an equation of state for the fluid is introduced
which is obtained from the 
Scherk-Schwarz compactification of a conformal field theory. The 
thermodynamic variables for each phase are then calculated.
In Sec.~\ref{sec:phase}, the thermodynamic phase structures are obtained. 
The final section is devoted to a summary and discussion. A technical issue about 
numerical integrations is treated in Appendix~\ref{sec:numerics}.
A justification of numerical results and a natural interpretation of phase diagram
are described in Appendix~\ref{sec:justification}.

\subsection*{Note Added}
During the preparation of this paper, we were informed that
a similar work had been completed independently~\cite{Marco}.
It is interesting to compare their results with ours.

\section{Hydrostatics of Axisymmetric Plasma Lumps}
\label{sec:hydro}

\subsection{Relativistic Navier-Stokes Equations}
\label{sec:NS}

The equation of motion for the boundary fluid is simply given by the conservation of the stress tensor~\cite{fluid-gravity2},
\be
	\nabla_\mu T^{\mu\nu}
	=
	0.
\label{energy_momentum_cons}
\ee
The stress tensor consists of the 
perfect fluid part, the dissipative part, and the surface contribution in a 
certain long wavelength limit as
\be
	T^{\mu\nu}
	=
	T^{\mu\nu}_{\mathrm{perfect}}
	+
	T^{\mu\nu}_{\mathrm{surface}}
	+
	T^{\mu\nu}_{\mathrm{dissipative}},
\ee
where the perfect fluid part takes the usual form as
\be
&&
	T^{\mu\nu}_{\mathrm{perfect}}
	=
	( \rho + P ) u^\mu u^\mu
	+
	P g^{\mu\nu}
\ee
with the pressure $P$, proper energy density $\rho$, and velocity field 
$u^\mu$.
The surface contribution can be written as
\be
	T^{\mu\nu}_{\mathrm{surface}}
	=
	\sigma ( n^\mu n^\nu - g^{\mu\nu} )
	\sqrt{ \pd \Phi \cdot \pd \Phi } \; \delta (\Phi)
\ee
with $\sigma$ being
a tension of the boundary.
We have assumed that the surface 
of the fluid is given by $\Phi(x^\mu)=0$ and  the unit 
normal of this surface is 
denoted by $n_{\mu} := \pd_\mu \Phi ( \pd \Phi \cdot \pd \Phi 
)^{-1/2}$. For static fluids which we will study in this paper one 
can show the dissipative term does not contribute to the equation of motion.

\begin{figure}[t!]
\begin{center}
\includegraphics[width=6cm]{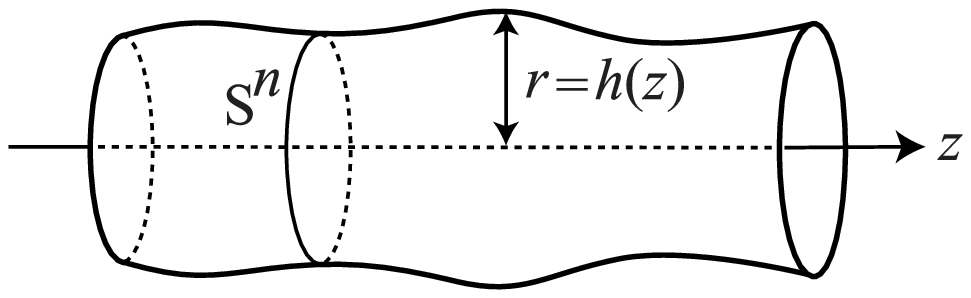}
\caption{{\textsf{An axisymmetric static equilibrium state of fluid in a $d=(n+3)$-dimensional flat spacetime schematically embedded in a three 
dimensional space.}}}
\label{fg:rz}
	\end{center}
\end{figure}

Now, we consider a
$d=(n+3)$-dimensional flat spacetime ($n \geq 1$)\footnote{
In fact, the conformal boundary, where the plasma fluid resides,
is $(d+1)$-dimensional if we take into account the existence of the Scherk-Schwarz circle, 
in which direction we assume the fluid configuration is uniform. Therefore, 
the total spacetime dimension of the AdS bulk is $d+2$.} in cylindrical coordinates
$x^\mu=(t,z,r,\phi^i)$, $(i=1,2,\cdots,n)$,
\be
	g_{\mu\nu} \dd x^\mu \dd x^\nu
 	&=&
	- \dd t^2 + \dd z^2 + \dd r^2 + r^2 \gamma_{ij}(\phi) \dd \phi^i 
          \dd \phi^j
\nonumber
\\
	&=&
	\eta_{ab} \dd x^{a} \dd x^b + r^2 \gamma_{ij}(\phi) \dd \phi^i 
        \dd \phi^j,
\ee
where $x^a=(t,z,r)$, $\eta_{ab} = \mathrm{diag}.(-1,1,1)$ and $ \gamma_{ij} 
\dd \phi^i \dd \phi^j$ is the line element of the unit $n$-sphere.
Non-zero components of the Christoffel symbol are
\be
	\Gamma^{a}_{\;\;ij}
	=
	- r \delta^a_r \gamma_{ij},
\;\;\;\;\;
	\Gamma^i_{\;\;ja}
	=
	r^{-1} \delta_a^r \delta^i_j,
\;\;\;\;\;
	\Gamma^{i}_{\;\;jk}
	=
	\bar{\Gamma}^i_{\;\;jk},
\ee
where the bar is used for the  variables with respect to the metric 
$\gamma_{ij}$. Assuming that the fluid is static in this frame ($u^\mu = 
\delta^\mu_t$), the perfect fluid part of the stress tensor and its divergence
 are given by
\be
	T^{\mu\nu}_{\mathrm{perfect}}
	=
	\left(
	\begin{array}{cccc}
		\rho	&	0	&	0	& 0	\\
		0	&	P	&	0	&	0 	\\
		0	&	0	&	P	&	0 	\\
		0	&	0	&	0	&	r^{-2} P 
\gamma^{ij} 
	\end{array}
	\right),
\;\;\;
	\nabla_\mu T^{\mu \nu}_{\mathrm{perfect}}
	=
	\left( 0, \frac{\pd P}{\pd z}, \frac{ \pd P }{ \pd r } , 0 \right),
\ee
where we have used an identity $ \bar{\nabla}_{i} \gamma_{jk} \equiv 0 $. The 
boundary surface of an axisymmetric fluid configuration can be given by 
$ \Phi(z,r) := r - h(z) = 0 $, where $h(z)$ is a height function (see 
Fig.~\ref{fg:rz}) assumed to be a single-valued function.
 Then, noting that the unit normal vector $n^\mu$ is given by
\be
	n^{\mu}
	=
	\frac{1}{ \sqrt{ 1+h^{\prime 2} }}
	\left(
		\delta^\mu_r - h^{\prime} \delta^\mu_z
	\right),
\;\;\;\;\;
	h^\prime := \pd_z h,
\ee
the surface contribution to the stress tensor and its divergence are given by
\be
&&
	T^{\mu \nu}_{\mathrm{surface}}
	=
	\frac{ \sigma \delta (r-h) }{ \sqrt{ 1+h^{\prime 2} } }
	\left(
		\begin{array}{cccc}
			1+h^{\prime 2}	&	0	&	0	& 0	\\
			0	&	-1	&	-h^\prime	&	0 	\\
			0	&	-h^\prime	&	-h^{\prime 2}	&	0 	\\
			0	&	0	&	0	&	- r^{-2} (1+h^{\prime 2}) \gamma^{ij} 
		\end{array}
	\right),
\nonumber
\\
&&
	\nabla_\mu T^{\mu \nu}_{\mathrm{surface}}
	=
	\sigma \delta (r-h)
	\Big( 0, - (n+1) h^\prime H, (n+1)H , 0 \Big).
\ee
Here, $H$ is the \textit{mean curvature} of the surface defined by
\be
	H(z)
	:=
	\frac{ 1 }{ n+1 }
	\left(
		- \frac{ h^{\prime\prime} }{ ( 1 + h^{\prime 2} )^{3/2} }
		+ \frac{ n }{ h \sqrt{ 1 + h^{\prime 2}  } }
	\right).
\ee
The first and second terms in the parenthesis correspond to the axial and 
azimuthal principal curvatures of the surface, respectively
 (see, e.g., \cite{Miyamoto}).

The non-trivial components of Eq. (\ref{energy_momentum_cons}) are
\be
&&
	\frac{ \pd P }{ \pd z }
	-
	(n+1) \sigma \delta (r-h) h^\prime H
	=
	0,
\nonumber
\\
&&
	\frac{ \pd P }{ \pd r }
	+
	(n+1) \sigma \delta (r-h) H
	=
	0.
\ee
Away from the boundary $r=h(z)$, we have
the equation in the bulk, which gives
\be
	P(z,r) = \mathrm{const}.
\ee
Integrating the equations of motion across the boundary, we have
\be
	P_> - P_{<} + (n+1) \sigma H (z) = 0,
\label{eq:balance}
\ee
where $P_{>}$ and $P_{<}$ are the pressures just outside and inside the 
boundary, respectively.
Now, we consider the case in which the pressure outside vanishes ($P_{>}=0$) 
for simplicity. Then, Eq.~(\ref{eq:balance}) reads
\be
	\frac{1}{n+1}
	\left(
	- \frac{ h^{\prime\prime} }{ ( 1 + h^{\prime 2} )^{3/2} }
	+ \frac{ n }{ h \sqrt{ 1 + h^{\prime 2}  } }
	\right)
	=
	\frac{P_{<}}{ (n+1)\sigma }
	=:
	H_0.
\label{eq:eom}
\ee
This is the governing equation that determines the axisymmetric equilibrium 
states of the fluid. This equation implies that the fluid surface is given by 
a \textit{constant mean curvature surface}. As we will see, this equation is 
derived by a variational principle
 which minimizes the surface area of the 
fluid while keeping the volume fixed.
A uniform tube found as a trivial solution of Eq.~(\ref{eq:eom}) will suffer from the Rayleigh-Plateau 
instability\footnote{The axisymmetric constant mean curvature surfaces 
in general dimensions were obtained in~\cite{MiyamotoMaeda}, and their 
geometric properties were investigated to compare those of black strings.}.

\subsection{The Equivalent Plateau Problem and \RP Instability}
\label{sec:RP}

In general, the interior volume and surface area of a fluid body are 
written as
\be
	V = \int_{\mathrm{interior}} \!\!\!\!\!\!\! \dd V,
\;\;\;
	A = \int_{\mathrm{surface}} \!\!\!\!\!\!\! \dd A.
\label{eq:general-av}
\ee
When we assume the staticity and axisymmetry of the fluid of which surface is 
given by $r=h(z)$,
the induced metric on the surface is given by 
\be
&&
	\dd s_{ \mathrm{surface} }^2
	=
	-\dd t^2 + ( 1+h^{\prime 2} ) \dd z^2 + h^2 \gamma_{ij}(\phi) 
        \dd \phi^i \dd \phi^j.
\ee
Now, we assume that the $z$-direction is compactified on a circle, $z 
\in [-L/2,L/2]$\footnote{This compactification corresponds to compactifying a 
spatial 
direction of the field theory coordinates in the Poincar\'{e} patch from the 
AdS point of view. However, it does not introduce the problem of singularities 
in the asymptotic IR region since the IR is cutoff by the horizon or the shrinking of the 
Scherk-Schwarz circle. We would like to thank T.~Wiseman for 
discussion on this point.}.
Noting that the volume and surface elements in Eq.~(\ref{eq:general-av}) are
 given by $ \dd V = r^n \sqrt{ \gamma } \, 
\dd r \wedge \dd z \wedge \dd \phi^1 
\wedge \cdots \wedge \dd \phi^n$ and $\dd A = h^n \sqrt{( 1+h^{\prime 2} ) 
\gamma } \, \dd z \wedge \dd \phi^1 \wedge \cdots \wedge \dd \phi^n $, we have
\be
&&
	V[h]
	=
	\Omega_n
	\int_{-L/2}^{L/2} \! \dd z \; h^{n+1}(z),
\;\;\;
	A[h]
	=
	(n+1) \Omega_n
	\int_{-L/2}^{L/2} \! \dd z \sqrt{ 1+h^{\prime 2} } \; h^n(z),
\nonumber
\\
&&	
	\Omega_n
	:=
	\frac{1}{n+1}
	\int_{S^n} \sqrt{ \gamma } \; \dd \phi^1 \wedge \cdots 
        \wedge \dd \phi^{n}
	=
	\frac{ \pi^{(n+1)/2} }{ \Gamma[ (n+1)/2 +1 ] }.
\ee

The equation of motion for the configuration that minimizes (or extremalizes) 
the surface area for a given volume is obtained by varying the action
\be
	I[h]
	:=
	\sigma_0 A[h] - p_0 V[h].
\label{eq:action}
\ee
Here, the constant $\sigma_0$ corresponds to the surface tension, while the 
constant $p_0$ corresponds to the pressure, but it is just a Lagrange 
multiplier to mathematically ensure that the volume is fixed. The 
Euler-Lagrange equation obtained by variation ($\delta_h I =0$) is equivalent
 to Eq.~(\ref{eq:eom}) with the identification of $p_0\leftrightarrow P_<$ and
 $\sigma_0 \leftrightarrow \sigma$.

It is noted that by adding a term corresponding to the rotation energy to the 
action~(\ref{eq:action}), which is roughly given by $ - \omega J$ with an 
angular velocity $\omega$ and an angular momentum $J$, the equation of $h(z)$ 
for a rigidly rotating non-relativistic fluid is obtained. In the rotational 
case, the surface deviates from a constant mean curvature surface
\cite{Cardoso,plasmaring1,plasmaring2}
\footnote{See Ref.~\cite{Marco} for more general relations between the equation of motion~(\ref{eq:balance}), called the Young-Laplace equation, and variational principles. They showed that in a fully covariant way the Young-Laplace equation can be obtained by variational principle of either the entropy maximization or potential energy minimization. It was also shown that these two variational principles reduce to the area minimization with volume fixing in the static situation. Its special case corresponds to ours. Their general proofs nicely relate the Young-Laplace equation (derived from the Navier-Stokes equation), thermodynamic equilibrium condition (the entropy maximization), mechanical equilibrium condition (the potential energy minimization), geometrical condition (the area minimization). More importantly it also supports the original argument in~\cite{plasmaball} (and our thermodynamic consideration in Sec.~\ref{sec:phase}), that the thermodynamics of finite energy black holes in the Scherk-Schwarz compactified AdS is mapped holographically to that of deconfined plasma lumps held by surface tension. The authors thank an anonymous referee for letting us know this point.}

Equation~(\ref{eq:eom}) has two trivial solutions representing a uniform 
tube (UT) and a spherical ball (SB),
\be
&
	h=h_{\mathrm{UT}} := r_0,
~~~~~~~~~~
\;\;\;\;\;
&
	H=H_{\UT}
	:=
	\frac{ n }{ (n+1) r_0 },
\nonumber
\\
&
	h=h_{\mathrm{SB}} := \sqrt{ R_0^2 - z^2 },
\;\;\;\;\;
&
	H=H_{\SB}
	:=
	\frac{ 1 }{ R_0 }.
\ee
As is well known, 
a translationally invariant cylindrical body is unstable if the 
linear dimension is longer than its circumference in $d=4$ (the \RP
 instability). Its onset mode in general dimensions is obtained by the 
following static perturbation of the uniform tube
\cite{Cardoso,MiyamotoMaeda}. First, we expand $h(z)$ around the uniform 
tube,
\be
	h(z)
	=
	r_0 + \varepsilon h_1(z) + O(\varepsilon^2).
\ee
Substituting this expansion into Eq.~(\ref{eq:eom}), we have the linear
perturbation equation at $ O(\varepsilon)$,
\be
	h_1^{\prime\prime} + \frac{ (n+1)^2 H_{\UT}^2 }{ n } h_1 = 0.
\ee
With a boundary condition, say $h_1^\prime(0)=0$, we have
\be
	h_1(z)
	=
	h^{(1)} \cos \left( k_{\mathrm{RP}} z \right),
\;\;\;\;\;
	k_{\mathrm{RP}}
	:=
	\frac{ \sqrt{ n } }{ r_0 },
\label{eq:linear-sol}
\ee
where $h^{(1)}$ is an integration constant.
Solution (\ref{eq:linear-sol}) corresponds to the marginally stable mode of 
the Rayleigh-Plateau instability. The uniform tube is unstable if the length 
of the cylinder $L$ satisfies $L > L_{\mathrm{RP}}:= 2\pi/k_{\mathrm{RP}}$. In 
other words, the uniform tube is unstable if the radius $r_0$ satisfies $r_0 
< r_{\mathrm{RP}} := \sqrt{n}L/2\pi$ for a given $L$. This dimensional 
dependence of the critical mode is quite similar to that of the \GL 
instability.
See Refs.~\cite{Cardoso,MiyamotoMaeda,Miyamoto} for more on the similarities between the \RP 
and \GL instabilities.

\section{Equation of State and Thermodynamic  Variables}
\label{sec:thermo}

In order to investigate the
thermodynamic properties of the fluid lumps in the 
context of AdS/CFT, we introduce an equation of state for the plasma fluid.
Then, the thermodynamic variables for each phase of fluid lumps are calculated.

\subsection{Equation of State}
\label{sec:eos}

The equation of state for the plasma is obtained by the Scherk-Schwarz 
compactification of a $(d+1)$-dimensional conformal field
theory~\cite{plasmaball}. 
The free energy in terms of temperature $T$ and volume $\mathcal{V}$ is given 
by
\be
	\mathcal{F}
	=
	\left( \rho_0 - \alpha T^{d+1} \right) \mathcal{V},
\ee
where $\rho_0$ is a vacuum energy and $\alpha$ is
some constant.
The pressure $P$, entropy $\mathcal{S}$, and energy $\mathcal{E}$ are given by
\be
&&
	P
	=
	- \left( \frac{ \pd \mathcal{F} }{ \pd \mathcal{V} } \right)_{T}
	=
	- \rho_0 + \alpha T^{d+1},
\nonumber
\\
&&
	\mathcal{S}
	=
	- \left( \frac{ \pd \mathcal{F} }{ \pd T } \right)_{ \mathcal{V} }
	=
	( d+1 ) \alpha \mathcal{V} T^d,
\nonumber
\\
&&
	\mathcal{E}
	:=
	\mathcal{F} + T\mathcal{S}
	=
	\left( \rho_0 + d\alpha T^{ d+1 } \right) \mathcal{V}.
\ee
We then introduce the energy density
$\rho$ and entropy density $s$, given as
\be
&&
	\rho
	:=
	\frac{ \mathcal{E} }{ \mathcal{V} }
	=
	\rho_0 + d\alpha T^{ d+1 },
\nonumber
\\
&&
	s
	:=
	\frac{\mathcal{S}}{ \mathcal{V} }
	=
	(d+1)\alpha^{1/(d+1)}( \rho_0 + P )^{d/(d+1)}.
\ee

From the parameters $\rho_0$, $\sigma$, and $\alpha$,
a characteristic length scale (whose existence violates the conformal 
invariance),
temperature, and entropy density can be defined as
\be
	l_0
	:=
	\frac{ \sigma }{ \rho_0 },
\;\;\;
	T_c
	:=
	\left( \frac{ \rho_0 }{ \alpha } \right)^{1/(d+1)},
\;\;\;
	s_0
	:=
	\left( \alpha \rho_0^d \right)^{1/(d+1)}.
\label{eq:chara}
\ee
The temperature in Eq.~(\ref{eq:chara}) is nothing but the critical temperature
of the confine-deconfine phase 
transition which occurs at $f:=\mathcal{F}/\mathcal{V}=-P=0$.

Now, we consider the thermodynamic
variables for the plasma lumps, that is,
not only those for 
their constituent plasma
but also the contributions from the surface term.
Their energy density and entropy density are given by
\be
&&
	T^{tt}
	=
	\rho
	+
	\sigma \delta (r-h) \sqrt{ 1+h^{\prime 2} },
\nonumber
\\
&&
	s
	=
	(d+1) s_0 \left( \frac{ T }{ T_c } \right)^{d}.
\ee
The energy, entropy, and Helmholtz free energy for the plasma lumps can be obtained by 
integrating the above densities,
\be
&&
	E
	=
	\int \left( T^{tt}_{\mathrm{perfect}} + T^{tt}_{\mathrm{surface}} 
        \right) \dd V
	=
	\rho V + \sigma A,
\nonumber
\\
&&
	S
	=
	\int s \dd V
	=
	s V,
\nonumber
\\
&&
	F=E-TS
	=
	-PV + \sigma A,
\label{eq:t-g}
\ee
where we have used that the energy density and entropy density are constant in the 
present system.

From the relations of $P=-\rho_0+\alpha T^{n+4}$ and $P = (n+1)\sigma H$ (note that $d=n+3$), we 
obtain a relation between the temperature $T$ and mean curvature $H$,
\be
	T
	=
	T_c \left[ 1+(n+1) \tilde{l}_0 HL \right]^{1/(n+4)},
\;\;\;
	\tilde{l}_0
	:=
	\frac{ l_0 }{ L }.
\label{eq:temperature-curvature}
\ee
We will discuss later a physically reasonable choice of
the dimensionless parameter $\tilde{l}_0$. 

It is convenient to define the following normalized 
dimensionless variables
when we draw phase diagrams,
\be
	\hat{E}
	:=
	\frac{ E }{ E_{\rp} },
\;\;\;
	\hat{S}
	:=
	\frac{S }{ S_{\SB} },
\;\;\;
	\hat{F}
	:=
	\frac{ F }{ F_{\SB} },
\;\;\;
	\hat{T}
	:=
	\frac{ T }{ T_{\rp} }
\,,
\label{eq:hat-variables}
\ee
where $E_{\rp}$ and $T_{\rp}$ are the energy and the temperature of the
critical uniform tube, and $S_{\SB}$ and $F_{\SB}$ are the entropy and free energy
of the spherical ball, respectively, which are given in the following subsection.

\subsection{Uniform Tube and Spherical Ball Phases}
\label{sec:UT-SB}

For the uniform tube, $h_{\UT}(z)=r_0$ with the length $L$, we can obtain 
the thermodynamic variables as functions of single parameter $r_0/L$,
\be
&&
	E_{\UT}
	=\rho_0
	\frac{ \Omega_n L^{n+2}}{ n }
	\left[
		(n^2+4n+1) \tilde{T}_{\UT}^{n+4} - 1
	\right]
	\left( \frac{ r_0 }{ L } \right)^{n+1},
\;\;\;
	S_{\UT}
	=
	s_0(n+4) \Omega_n L^{n+2} \tilde{T}_{\UT}^{n+3}
	\left( \frac{ r_0 }{ L } \right)^{n+1},
\nonumber
\\
&&
	F_{\UT}
	=\rho_0
	\frac{\Omega_nL^{n+2}}{n} \left( \tilde{T}^{n+4}_{\UT} - 1 \right)
	\left( \frac{ r_0 }{ L } \right)^{n+1},
\;\;\;
	T_{\UT}
	=T_c
	\left[
		1 + n \tilde{l}_0 \left( \frac{L}{r_0} \right) 
	\right]^{1/(n+4)},
\label{TDVUS}
\ee
where $\tilde{T}_{\UT}:=T_{\UT}/T_c$.
The critical values for the $\rp$ instability are obtained just by
setting $r_0 = r_\rp$ $(=\sqrt{n}L/2\pi)$ in Eq.~(\ref{TDVUS}).
For instance, the energy and temperature of the
critical tube are given by
\be
	E_{\rp}
	=\rho_0
	\frac{ \Omega_n L^{n+2}}{ n }
	\left( \frac{ \sqrt{n} }{ 2\pi } \right)^{n+1}
	\left[
		( n^2 + 4n + 1 ) \tilde{T}_{\rp}^{n+4} - 1
	\right],
\;\;\;\;
	T_{\rp}
	=
	T_c
	\left(
		1 + 2\pi \sqrt{n} \; \tilde{l}_0
	\right)^{1/(n+4)},
\label{eq:tilde-RP}
\ee
where $ \tilde{T}_{\rp}:=T_\rp/T_c $.

For the spherical ball, $h_{\SB}(z)=\sqrt{R_0^2-z^2}$ in 
the period $L$, we have the thermodynamic variables as functions of 
a single parameter $R_0/L$,
\be
&&
	E_{\SB}
	=\rho_0
	\frac{ \Omega_{n+1}L^{n+2} }{ n+1 }
	\left[
		( n^2+5n+5 ) \tilde{T}_{\SB}^{n+4} - 1
	\right]
	\left( \frac{ R_0 }{ L } \right)^{n+2},
\nonumber
\\
&&
	S_{\SB}
	=s_0
	(n+4)\Omega_{n+1}L^{n+2} \tilde{T}_{\SB}^{n+3}
	\left( \frac{ R_0 }{ L } \right)^{n+2},
\nonumber
\\
&&
	F_{\SB}
	=\rho_0
	\frac{ \Omega_{n+1} L^{n+2}}{ n+1 }
	\left( \tilde{T}_{\SB}^{n+4} - 1 \right)
	\left( \frac{ R_0 }{ L } \right)^{ n+2 },
\;\;\;
	T_{\SB}
	=T_c
	\left[
		1 + (n+1) \tilde{l}_0 \left( \frac{L}{R_0} \right)
	\right]^{1/(n+4)},
\ee
where $\tilde{T}_{\SB} := T_{\SB}/T_c$.

According to the definition~(\ref{eq:hat-variables}), we have the
following dimensionless quantities,
\be
&&
	\hat{E}_{\UT}
	=
	\left( \frac{ 2\pi }{ \sqrt{n} } \right)^{n+1}
	\frac{
		( n^2+4n+1 ) [ 1 + n \tilde{l}_0 (L/r_0) ] - 1
	}
	{
		( n^2+4n+1 )( 1+2\pi\sqrt{n} \; \tilde{l}_0 ) - 1
	}
	\left( \frac{ r_0 }{ L } \right)^{n+1},
\nonumber
\\
&&
	\hat{S}_{\UT}
	=
	\frac{ \Omega_n }{ \Omega_{n+1} }
	\left[
		\frac{ 1+n\tilde{l}_0 (L/r_0) }{ 1+(n+1)\tilde{l}_0(L/R_0) }
	\right]^{(n+3)/(n+4)}
	\left( \frac{ r_0 }{ L } \right)^{n+1}
	\left( \frac{ L }{ R_0 } \right)^{n+2},
\nonumber
\\
&&
	\hat{F}_{\UT}
	=
	\frac{ n^n \Omega_n }{ (n+1)^{n+1}\Omega_{n+1} }
	\frac{ \tilde{T}_\UT^{n+4} - 1 }{ \tilde{l}_0 },
\;\;\;
	\hat{T}_{\UT}
	=
	\left[
		\frac{ 1 + n \tilde{l}_0 (L /r_0) }{ 1 + 2\pi \sqrt{n} \; 
	\tilde{l}_0 }
	\right]^{1/(n+4)}
\label{eq:hat-UT}
\ee
for the uniform tube phase, and
\be
&&
	\hat{E}_{\SB}
	=
	\frac{ n\Omega_{n+1} }{ (n+1)\Omega_n }
	\left( \frac{ 2\pi }{ \sqrt{n} } \right)^{n+1}
	\frac{
		( n^2+5n+5 )[ 1+(n+1) \tilde{l}_0 ( L / R_0 ) ] - 1
	}
	{
		( n^2+4n+1 )( 1 + 2\pi\sqrt{n} \; \tilde{l}_0 ) - 1
	}
	\left( \frac{ R_0 }{ L } \right)^{n+2},
\nonumber
\\
&&
	\hat{T}_{\SB}
	=
	\left[
		\frac{ 1+(n+1) \tilde{l}_0 ( L/R_0 ) }{ 1 + 2\pi\sqrt{n}\;
\tilde{l}_0 }
	\right]^{1/(n+4)},
\;\;\;
	\hat{S}_{\SB}
	=
	\hat{F}_{\SB}
	\equiv
	1
\label{eq:hat-SB}
\ee
for the spherical ball phase.

Now, for each value of $\tilde{l}_0$,
which should be specified,
we obtain thermodynamic relations such as
$\hat{S}=\hat{S}(\hat{E})$ and 
$\hat{F}=\hat{F}(\hat{T})$  
implicitly via the parameter $r_0/L \in [0,+\infty)$ for the 
uniform tube phase and via $R_0/L \in [0,1/2]$ for the spherical ball phase, 
respectively\footnote{Because we normalize the entropy by that of the 
spherical ball, Eq.~(\ref{eq:hat-variables}), the expression of $\hat{S}_{\UT}$ in
Eq.~(\ref{eq:hat-UT}) contains the radius of the spherical ball $R_0$. When one 
draws the ($\hat{E}_\UT$,$\hat{S}_{\UT}$) curve by varying the parameter $r_0$, 
this $R_0$ should be regarded as a function of $r_0$. Such a relation 
$R_0=R_0(r_0)$ is obtained by the combination of $R_0=R_0(\hat{E})$ obtained 
by inversing the relation in Eq.~(\ref{eq:hat-SB}) and 
$\hat{E}=\hat{E}(r_0)$ in Eq.~(\ref{eq:hat-UT}). Similar transformations are 
needed to draw the ($\hat{E}_\NUT$,$\hat{S}_{\NUT}$) curve [see $\hat{S}_{\NUT}$ 
in Eq.~(\ref{eq:hat-NUT})].\label{fn:R0-dep}}.
Note that due to the fact $R_0\leq L/2$ (i.e., the condition that a ball fits the period), the spherical ball phase has an upper 
bound of energy and a lower bound of the temperature.

\begin{center}
\begin{figure*}[b!]
\setlength{\tabcolsep}{ 40 pt }
		\begin{tabular}{cc}
			(a) & (b) \\
			\includegraphics[width=6cm]{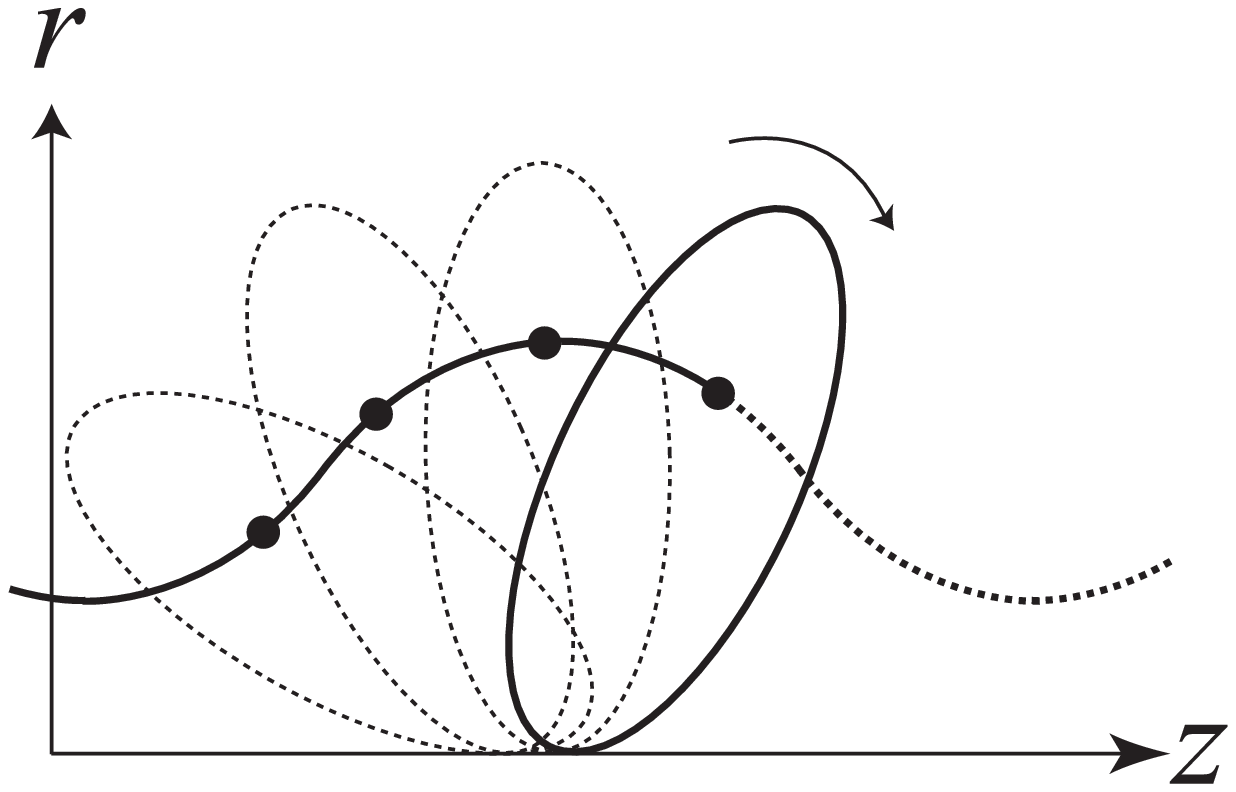} &
			\includegraphics[width=6cm]{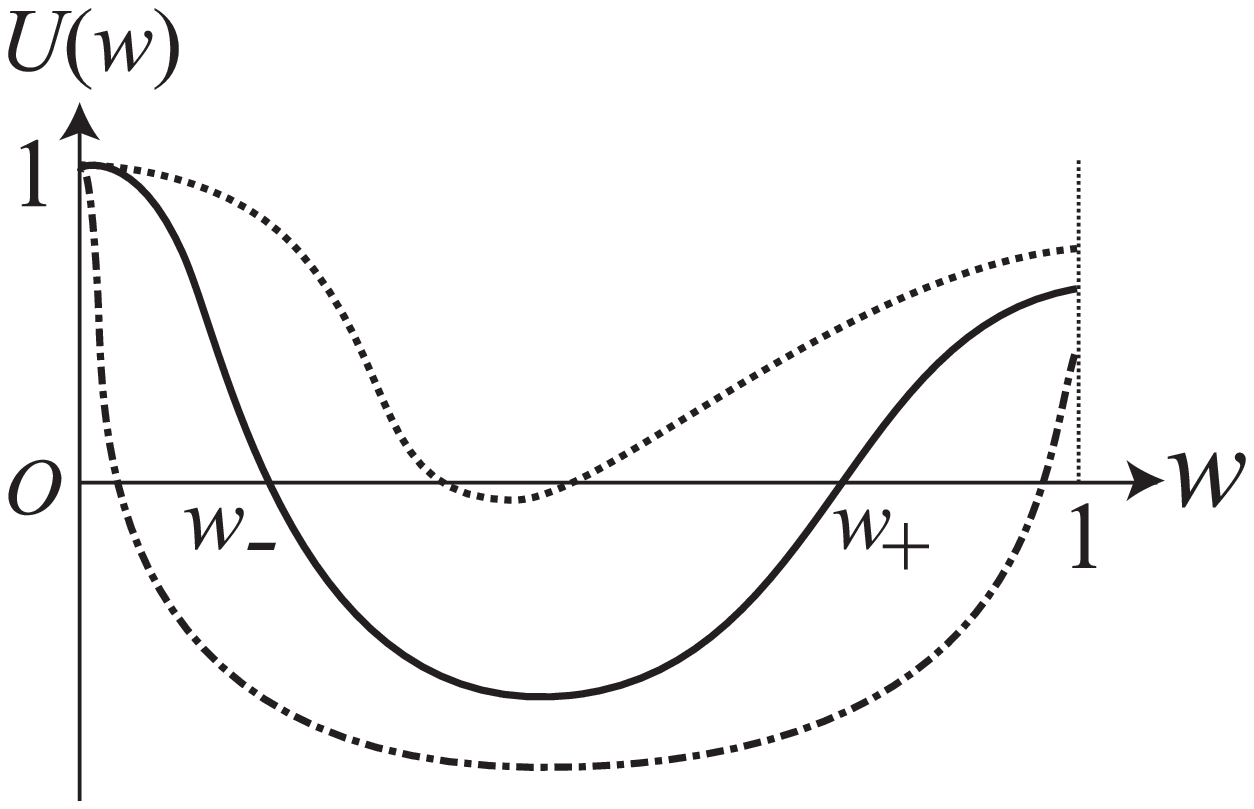}\\
		\end{tabular}
\caption{\textsf{(a) The non-uniform tube in 
$d=4$ is called Delaunay's unduloid in geometry, which is the surface of revolution of an 
elliptic catenary. The elliptic catenary is obtained by rotating an ellipse 
along a line (i.e., the $z$-axis) and tracing the focus, marked by small solid circles.
(b) A schematic picture of the potential $U(w)$, defined in 
Eq.~(\ref{eq:potential-form}), for three values of $\lambda=w_-/w_+$. From the 
top curve to the bottom, the value of $\lambda$ decreases within the range 
$0<\lambda<1$.
}}
\label{fg:unduloid}
\end{figure*}
\end{center}

\subsection{Non-uniform Tube Phase}
\label{sec:NUT}

As given in the preceding sections, Eq.~(\ref{eq:eom}) 
is solved by the uniform tube and the spherical ball.
In addition to these two solutions, 
Eq.~(\ref{eq:eom}) allows the third solution, which is a non-trivial 
constant mean curvature surface corresponding to the non-uniform tube (NUT)~\cite{MiyamotoMaeda}. 
For the $d=4$ case, this solution is known as \textit{Delaunay's unduloid}
\cite{unduloid}, which is the surface of revolution of an elliptic catenary 
[see Fig.~\ref{fg:unduloid}(a)]. Although a certain
 parametric representation of the curve is known for the $d=4$ case, 
for $d>4$, we have to 
integrate Eq.~(\ref{eq:eom}) numerically. As we will see below, if we 
introduce a `non-uniformness parameter' $\lambda:=r_-/r_+$, where $r_-$ and 
$r_+$ are the smallest and largest radii of a non-uniform tube, all 
non-uniform tube solutions are described by a one-parameter family of 
$\lambda$, ($0 < \lambda < 1$).

First, we introduce the following dimensionless variables,
\be
	y
	:=
	H_0 z,
\;\;\;\;\;
	w
	:=
	H_0 h(z),
\ee
where $H_0$ is the mean curvature of the non-uniform tube.
Then, the first integral of Eq.~(\ref{eq:eom}) is written in a potential 
form,
\be
	\left( \frac{\dd w}{\dd y} \right)^2 + U(w)
	=
	0\ ,
\;\;\;\;\;\;
	U(w)
	:=
	1 
	-
	\left(
		\frac{ w^{n} }{ w^{n+1} + K }
	\right)^2,
\label{eq:potential-form}
\ee
where $K$ is an integration constant.
We assume that there are two zero points of the 
potential $U(w)$ denoted by $w_+$ and $w_-$ ($0 < w_- < w_+$).
$w_+$ and $w_-$
correspond to the maximum and minimum (dimensionless) radii of the non-uniform
tube, respectively. Thus, $\lambda= w_-/w_+ \to 0$ 
and $\lambda \to 1$  correspond to the ball 
and the critical  tube,
respectively [see Fig.~\ref{fg:unduloid}(b) for a
schematic picture of $U(w)$]. 
If we first give a $\lambda$, from $U(w_\pm)=0$, we obtain
\be
	w_+
	=
	\frac{ 1 - \lambda^n }{ 1 - \lambda^{n+1} },
\;\;\;\;\;
	w_-
	=
	\frac{ \lambda (1 - \lambda^n) }{ 1 - \lambda^{n+1} },
\;\;\;\;\;
	K
	=
	\frac{ ( 1-\lambda ) \lambda^n ( 1-\lambda^n )^n }{
	( 1-\lambda^{n+1} )^{ n+1 } }.
\label{eq:Cw}
\ee
Using $\dd z = (\dd z/\dd h) \dd h = \dd w/( H_0 \sqrt{ -U } )$, one can 
express the period $L$, surface area $A$ and volume $V$ per the period in 
terms of $\lambda$ and $H_0$,
\be
&&
	L_{\mathrm{NUT}}
	=
	\frac{2}{H_0}
	\int_{w_-}^{w_+} \dd w\;
	\frac{ 1 }{ \sqrt{ -U(w) } }
	=:
	\frac{1}{H_0} \tilde{L}(\lambda)
	,
\nonumber
\\
&&
	A_{\mathrm{NUT}}
	=
	\frac{ 2(n+1)\Omega_n}{ H_0^{n+1} }
	\int_{w_-}^{w_+} \dd w\; w^n \sqrt{ \frac{1-U(w)}{-U(w)} }
	=:
	\frac{ (n+1)\Omega_n }{ H_0^{n+1} } \tilde{A}(\lambda),
\nonumber
\\
&&
	V_{\mathrm{NUT}}
	=
	\frac{ 2\Omega_n }{ H_0^{n+2} }
	\int_{w_-}^{w_+} \dd w\; \frac{ w^{n+1} }{ \sqrt{-U(w)} }
	=:
	\frac{ \Omega_n }{ H_0^{n+2} } \tilde{V}(\lambda).
\label{eq:NUB-quant}
\ee

As a result, for the non-uniform  tube, we can obtain
the thermodynamic dimensionless quantities as functions 
of the non-uniformness parameter $\lambda$,
\be
&&
	\hat{E}_{\NUT}
	=
	n
	\left( \frac{ 2\pi }{ \sqrt{n} } \right)^{n+1}
	\frac{
		[ (n+3) \tilde{V} + \tilde{A} ][ 1 + (n+1) \tilde{l}_0 
	\tilde{L}  ] + \tilde{V} - \tilde{A}
	}
	{
		[ ( n^2 + 4n + 1 )( 1 + 2\pi \sqrt{n} \; \tilde{l}_0 ) - 1 ]
 	\tilde{L}^{n+2}
	},
\nonumber
\\
&&
	\hat{S}_{\NUT}
	=
	\frac{ \Omega_n }{ \Omega_{n+1} }
	\left[
		\frac{ 1+(n+1)\tilde{l}_0 \tilde{L} }{ 1+(n+1)
	\tilde{l}_0(L/R_0) }
	\right]^{(n+3)/(n+4)}
	\frac{ \tilde{V} }{ \tilde{L}^{n+2} }
	\left( \frac{ L }{ R_0 } \right)^{n+2},
\nonumber
\\
&&
	\hat{F}_{\NUT}
	=
	\frac{ n+1 }{ n^n }
	\frac{ ( \tilde{T}^{n+4}_{\NUT} -1 )^n ( \tilde{A} - \tilde{V} ) }
		 { \tilde{l}_0^n \tilde{L}^{n+1} },
\nonumber
\\
&&
	\hat{T}_{\NUT}
	=
	\frac{ \Omega_n }{ (n+1)^{n+1} \Omega_{n+1} }
	\left( \tilde{T}_\NUT^{n+4} - 1 \right)^{n+2}
	\frac{ \tilde{A} - \tilde{V} }{ \tilde{l}_0^{n+2} \tilde{L}^{n+2} },
\label{eq:hat-NUT}
\ee
where $\tilde{T}_{\NUT}:=T_{\NUT}/T_c$, and $\tilde{L}(\lambda)$, 
$\tilde{A}(\lambda)$, and $\tilde{V}(\lambda)$ are 
the dimensionless variables defined in Eq.~(\ref{eq:NUB-quant}). 
For each value of $\tilde{l}_0$, which should be specified,
 we have thermodynamic relations, $\hat{S}=\hat{S}(\hat{E})$ and $\hat{F}=
\hat{F}(\hat{T})$, implicitly via the
non-uniformness parameter $\lambda \in (0,1)$, (see
 also Footnote~\ref{fn:R0-dep}).

Now, we have expressed all thermodynamic  variables as functions of the 
parameter $\lambda$ in the integration form.
Although the integrations in Eq.~(\ref{eq:NUB-quant}) are of course finite,
the integrands diverge at the both ends of the integration range
since $U(w_\pm)=0$. Therefore, one needs some manipulations for an accurate numerical 
integration, especially in the highly deformed regime $\lambda \ll 1$.
We present some technical prescriptions making it possible
to figure the fine structures of phase diagrams in Appendix~\ref{sec:numerics}.

\section{Phase Structure}
\label{sec:phase}

In the previous section, we have described the necessary 
thermodynamic variables of each phase by the appropriate parameters: the 
uniform  tube is parameterized by $r_0/L \in [0,\infty)$; the 
spherical ball  by $R_0/L \in [0,1/2]$; and the non-uniform  tube 
 by $\lambda \in (0,1)$.
To draw  phase diagrams, we have to specify the undetermined 
parameter $\tilde{l}_0=l_0/L=\sigma/(\rho_0 L)$.
Although we are treating the sharp boundary as a fluid surface, in reality 
the boundary surface has a thickness of
order  $T_c^{-1} \sim \sigma/\rho_0 = l_0$
\cite{plasmaball,plasmaring1}. Therefore, we obviously should work in the 
limit $l_0 \ll L$, where the thickness of the boundary can be ignored (except 
for highly deformed configurations). Otherwise higher-derivative 
contributions to the surface stress tensor would dominate. Hence, we take 
$\tilde{l}_0 = 1.0 \times 10^{-2}$ in all concrete examples hereafter.
 It should be
 stressed, however, that the qualitative aspects of the phase diagrams do not 
depend on a specific choices of parameter $\tilde{l}_0$. In particular, we 
have 
confirmed that the critical dimensions which we will find do not change for 
other choice of $\tilde{l}_0$ within the range of $1.0 \times 10^{-2} \le 
\tilde l_0 \lesssim 1$. It is noted that such a choice of small $\tilde{l}_0$ 
corresponds to $0 < T_{\rp}/T_c - 1 \ll 1$ as one can see from 
Eq.~(\ref{eq:tilde-RP}).
That is, we only consider the plasma just above the critical temperature $T_c$.

\subsection{Microcanonical Ensemble: $\hat{S}=\hat{S}(\hat{E})$}
\label{sec:microcan}

First we discuss the phase transitions among the plasma lumps in the microcanonical ensemble.
Hence, we consider the $\hat{E}$-$\hat{S}$ relation.
According to the behaviors of each phase curve, e.g.,
the number of cusps appearing in the $\hat{E}$-$\hat{S}$ diagram here,
we divide the dimensions into three groups, 
i.e., $4 \leq d \leq 9$, $10 \leq d \leq 12$, and $d \geq 13$.
We now discuss the possible phase transitions between the three phases 
(a spherical ball, and uniform and non-uniform tubes)
with the criterion that the maximum entropy state for a fixed energy is favored.

\subsubsection{$4 \leq d \leq 9$}
\label{sec:class1}

\begin{figure}[b!]
	\begin{center}
		\begin{tabular}{c}
			$d=5$  \\
			\includegraphics[width=5.5cm]{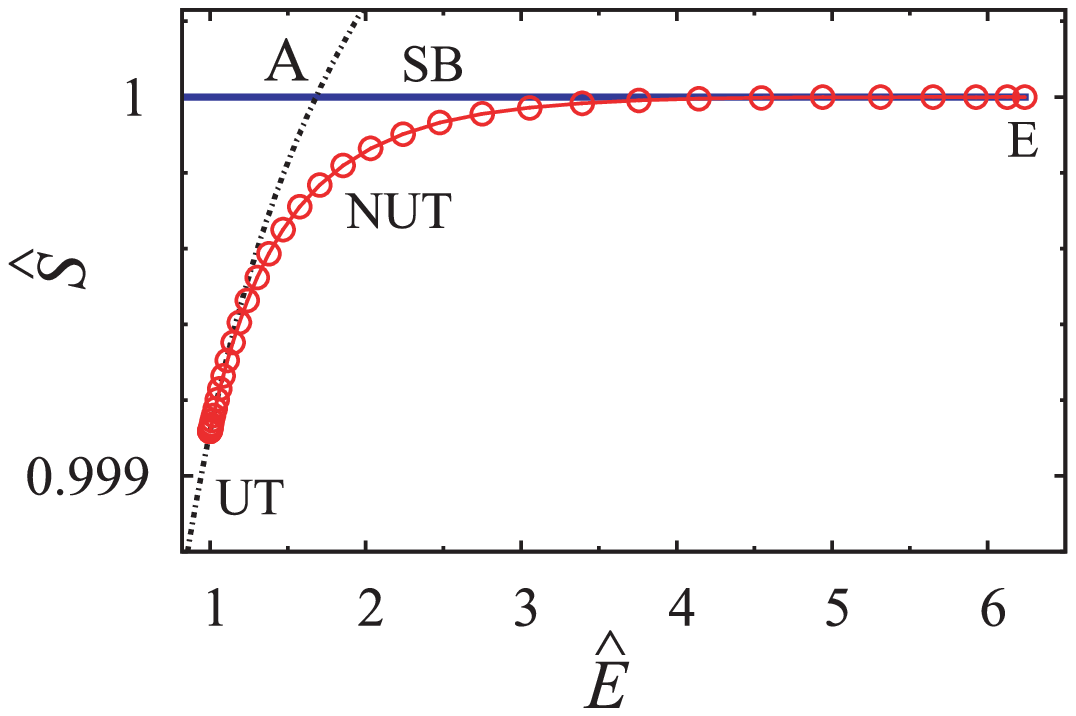}  \\
		\end{tabular}
\caption{\textsf{The energy-entropy diagram for $d=5$, containing the 
phases of the spherical ball (SB, blue-solid line), uniform tube (UT, 
black-dashed curve), and non-uniform tube (NUT, red-solid curve with 
small circles of data). The maximum entropy state for a given energy is favored.
There is no cusp on the non-uniform tube phase, which is the case for $4 \leq d \leq 9$.}}
\label{fg:5d(e,s)}
	\end{center}
\end{figure}

In this first class of dimensions, the branches of the
uniform  tube and spherical
 ball intersect at the point labeled by A in Fig.~\ref{fg:5d(e,s)}. 
The spherical ball always has a larger entropy than the critical tube.

From Fig.~\ref{fg:5d(e,s)}, one can see that the branch of non-uniform tubes
emerges from the \RP critical point, located at $\hat{E}=1$, and reaches 
the end of the spherical ball
 branch located at the point E, where the ball `touches itself' at the 
boundary $z=\pm L/2$. The most important point is that the non-uniform 
 tube branch 
always has a smaller entropy than the other two. 
Thus, the phase diagram suggests that the non-uniform branch is never favored. 
Suppose that we have a fat uniform tube (i.e., large $\hat{E}$) and decrease 
its energy $\hat{E}$. As $\hat{E}$ decreases, 
the uniform branch meets the spherical ball branch at the point A, 
and at this point the uniform  tube transits to a ball with 
a discontinuous jump in configuration. This transition is of first order.

Here, we can see that the energy-entropy diagram in $d=5$ is quite similar to 
the mass-entropy digram of the black hole-black string system
in the asymptotically locally flat 5 and 6 dimensional Kaluza-Klein space 
(see Fig.~6 in~\cite{KudohWiseman}). Only one apparent difference is that the
localized phase, i.e., the spherical plasma ball does not deform unlike a
localized caged black hole.

\subsubsection{$10 \leq d \leq 12$}
\label{sec:class2}

\begin{figure}[b!]
	\begin{center}
		\setlength{\tabcolsep}{ 2 pt }
		\begin{tabular}{ccc}
			(a) $d=10$ &
			(b) $d=11$ &
			(c) $d=12$  \\
			\includegraphics[width=5.5cm]{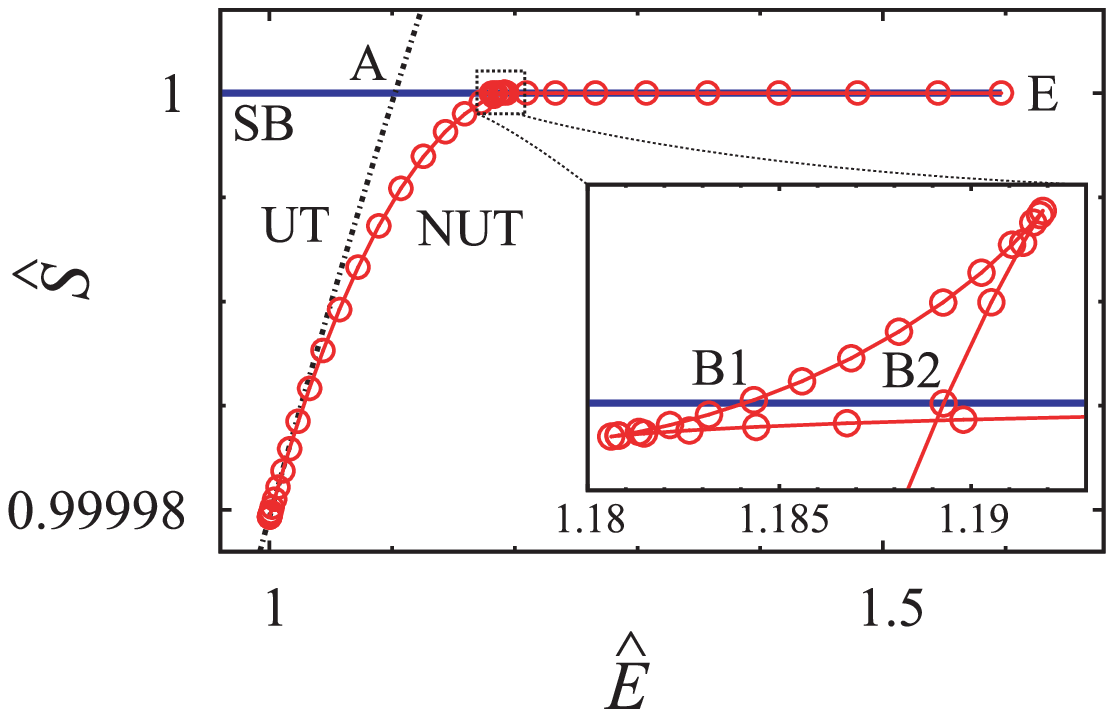} &
			\includegraphics[width=5.5cm]{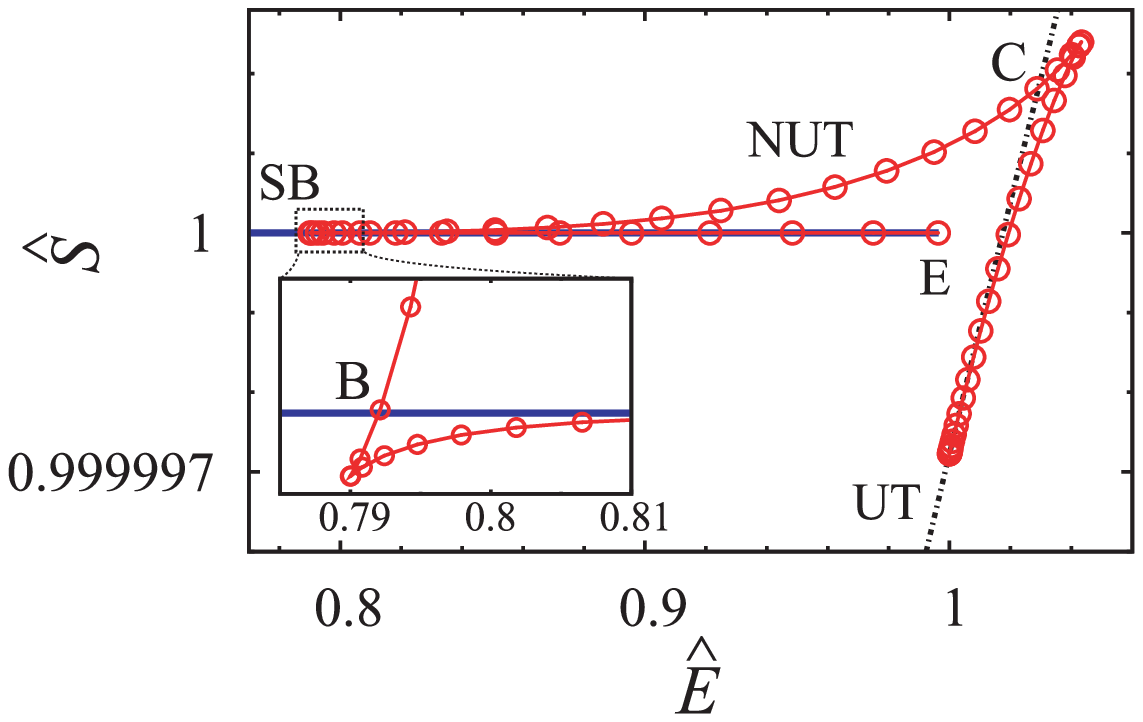} &
			\includegraphics[width=5.5cm]{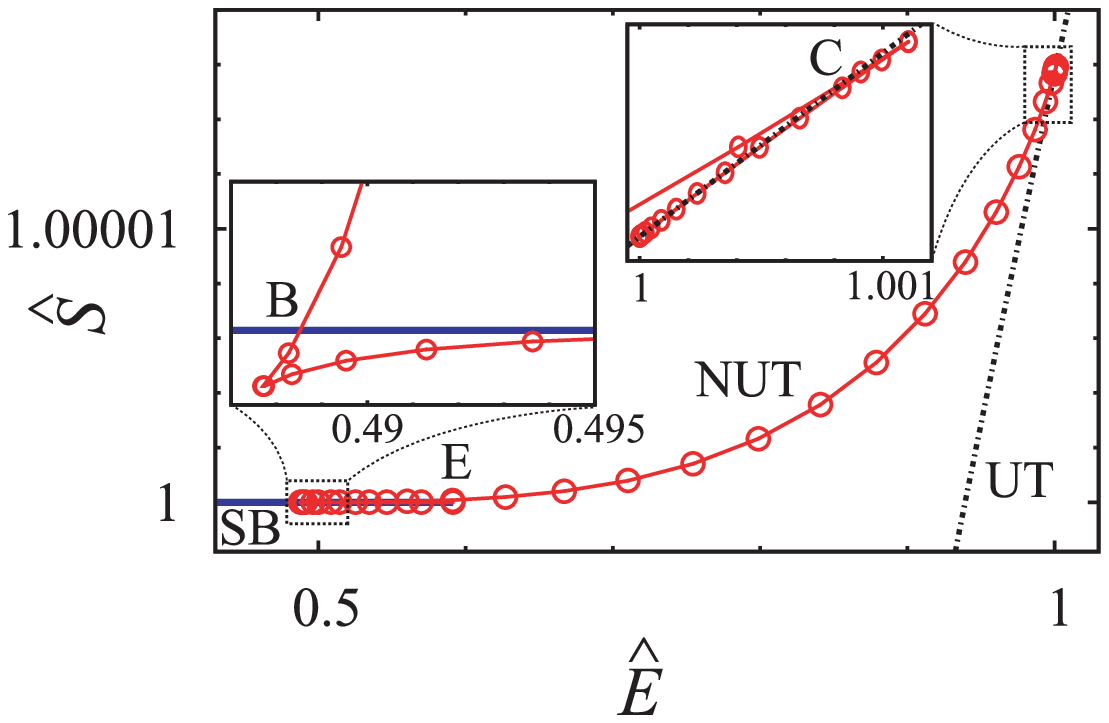}\\
		\end{tabular}
\caption{\textsf{The energy-entropy diagrams for (a) $d=10$, (b) $d=11$, and (c) $d=12$.
There appear two cusps on the non-uniform tube phase in this class of dimensions.}}
\label{fg:10d(e,s)}
	\end{center}
\end{figure}

In this class of dimensions, the phase diagrams are somewhat complicated due 
to the appearance of two cusps in the ($\hat{E},\hat{S}$) diagram. The 
($\hat{E},\hat{S}$) diagrams for $d=10,11$ and 12
are shown in Fig.~\ref{fg:10d(e,s)}.
The uniform  tube and spherical ball intersect in $d=10$ at the point A, 
while they do not intersect any more in $d=11$ and $12$ 
(in fact, ${}^\forall d \geq 11$). The spherical ball has a larger entropy
than the critical tube in $d=10$ and $11$, while this is not the 
case in $d=12$ (in fact, ${}^\forall d \geq 12$). As can be seen from 
these observations, there appear rich and non-trivial phase structures in 
these transient dimensions.

From the fact that the non-uniform branch has two cusps in the ($\hat{E},\hat{S}$) diagrams,
the non-uniform tube branch can be divided into three branches 
(i.e., three smooth curves): the first branch leaves the \RP critical point; 
the second one is connected to the end point E of the spherical ball branch; 
the third one is the curve between the first and the second ones.
 The first and third branches correspond to the non-uniform tube branch 
appearing in the lower dimensions ($4 \leq d \leq 9$). 
We shall first discuss the case of $d=11$.
The non-uniform tube emerges from the \RP critical point.
$\hat{E}$ increases at first,
however, it turns to decrease, and then to increase again.
Eventually the non-uniform tube branch 
 reaches the end point E of the spherical ball branch. 
During this change, the non-uniform tube branch intersects with the uniform 
tube branch at the point C and with the spherical ball branch
at the point B. 
Thus, if we begin with a fat uniform tube (i.e., large $\hat{E}$) and 
decreases $\hat{E}$, the uniform tube transits to the non-uniform tube at 
the point C with a discrete jump in configuration. As $\hat{E}$ decreases 
further, the non-uniform tube transits to a spherical  ball
 at the point B
 with a discrete jump in configuration.

The phase structures of $d=12$ is quite similar to that in $d=11$.
 That is, if we begin with a fat uniform 
 tube and decrease the energy, the favored phases
shift as (uniform tube) 
$\to$ (non-uniform tube) $\to$ (spherical ball).
These two phase transitions are accompanied by the discrete jumps
in configuration, and hence they are of first order. 
However,
the phase structure in $d=10$ is different from those of $d=11$ and 
$d=12$, although there are two cusps in all cases. The branches of 
uniform  tube and spherical ball still intersects (at the point A) .
The spherical ball has a larger entropy than the critical uniform  tube. 
As a consequence, the behavior of the non-uniform  tube branch is similar to 
those in $4 \leq d \leq 9$ except that two quite small cusps appear at the 
middle of the branch. That is, if we begin with a fat uniform  tube and 
decreases the energy, the uniform  tube transits to the spherical ball at 
the point A. The non-uniform  tube has nothing to do with this transition.

\subsubsection{$d \geq 13$}
\label{sec:class3}

\begin{figure}[t!]
	\begin{center}
		\setlength{\tabcolsep}{ 5 pt }
		\begin{tabular}{c}
			$d=13$  \\
			\includegraphics[width=5.5cm]{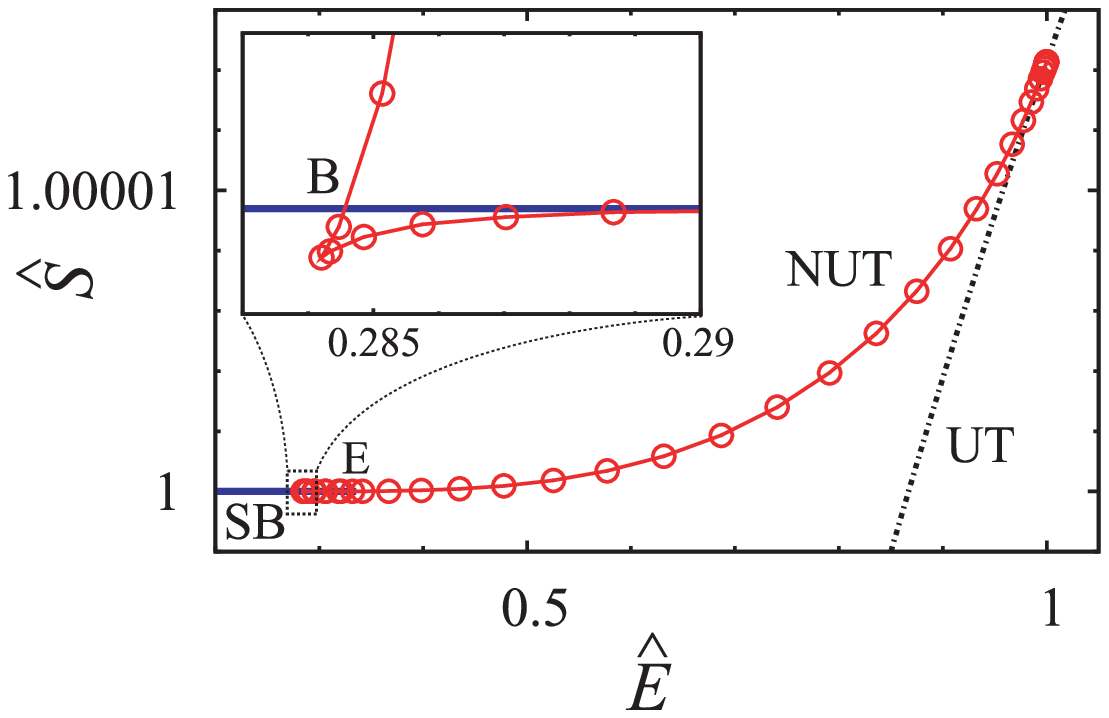}\\
		\end{tabular}
\caption{\textsf{The energy-entropy diagram for $d=13$.
There exists one cusp on the non-uniform tube phase, which is the case for $d \geq 13$.}}
\label{fg:13d(e,s)}
	\end{center}
\end{figure}

As mentioned before, the branches of the uniform tube and 
spherical ball do not intersect each other in this class of dimensions. 
The spherical ball phase has a smaller entropy than the critical uniform 
tube in this class of dimensions.
The cusp near the \RP critical point disappears at $d=13$, 
while the cusp near the spherical ball phase remains 
(we have confirmed this up to $d=15$). Therefore, 
if we begin with a fat uniform 
 tube (i.e., large $\hat{E}$) and decrease $\hat{E}$, the non-uniform branch 
meets the \RP critical point, and at this point it smoothly transits to the 
 non-uniform  tube, which has a larger entropy. This transition  
is not accompanied by a discrete jump in configuration, and therefore
is of second or higher order.
As $\hat{E}$ decreases further, the non-uniform branch intersects with 
the ball phase at the point B, and transits to a ball.
This transition is of first order.

Thus, the smooth transition from the uniform tube to the non-uniform tube is realized for $d \geq 13$. We conclude that the critical 
dimension in the microcanonical ensemble is $d_\ast^{\mathrm{microcan}}=13$. 
Thus, the critical dimension for the gravity dual is expected to be around
$D_\ast^{\mathrm{microcan}}=d_\ast^{\mathrm{microcan}}+2=15$\footnote{ 
Here, we adopt the definition of critical dimension by Sorkin~\cite{Sorkin}.
However, if we define a critical dimension, from a non-perturbative viewpoint, 
by the dimension at and above which the transition from a uniform tube 
(or uniform black string) to a non-uniform tube (or non-uniform black string) 
is realized, we find $ D_\ast^{\mathrm{microcan}} = 13 $.}.

\subsection{Canonical Ensemble: $\hat{F}=\hat{F}(\hat{T})$}
\label{sec:cano}

\begin{figure}[t!]
	\begin{center}
		\setlength{\tabcolsep}{ 3 pt }
		\begin{tabular}{ ccc }
			(a) $d=5$ &
			(b) $d=10$ &
			(c) $d=11$ \\
			\includegraphics[width=5.5cm]{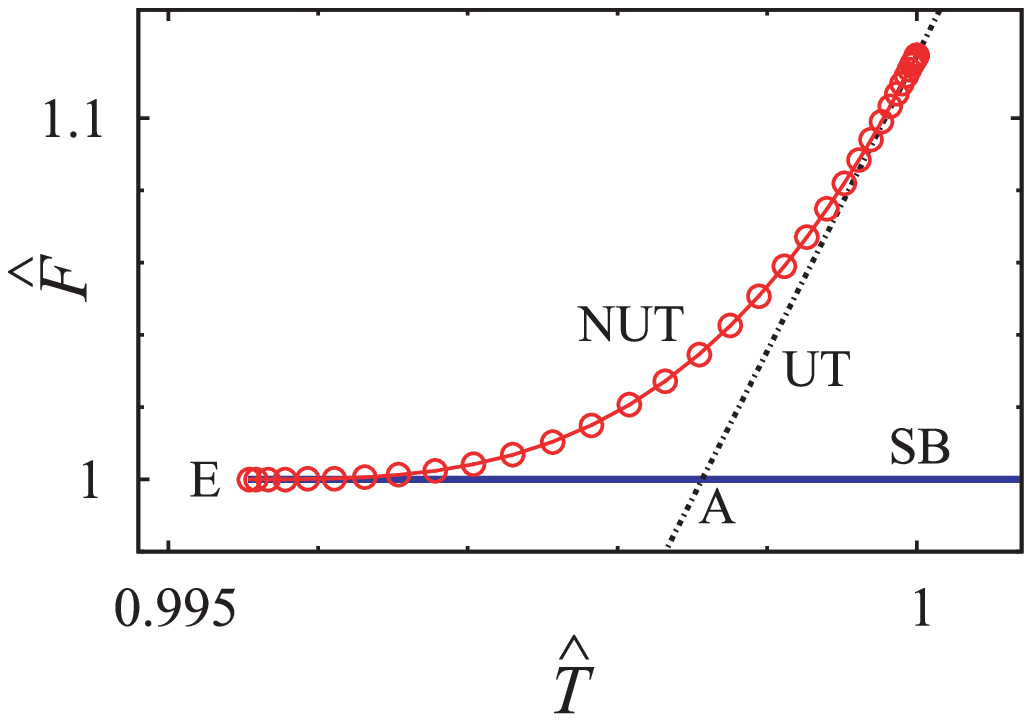}  &
			\includegraphics[width=5.5cm]{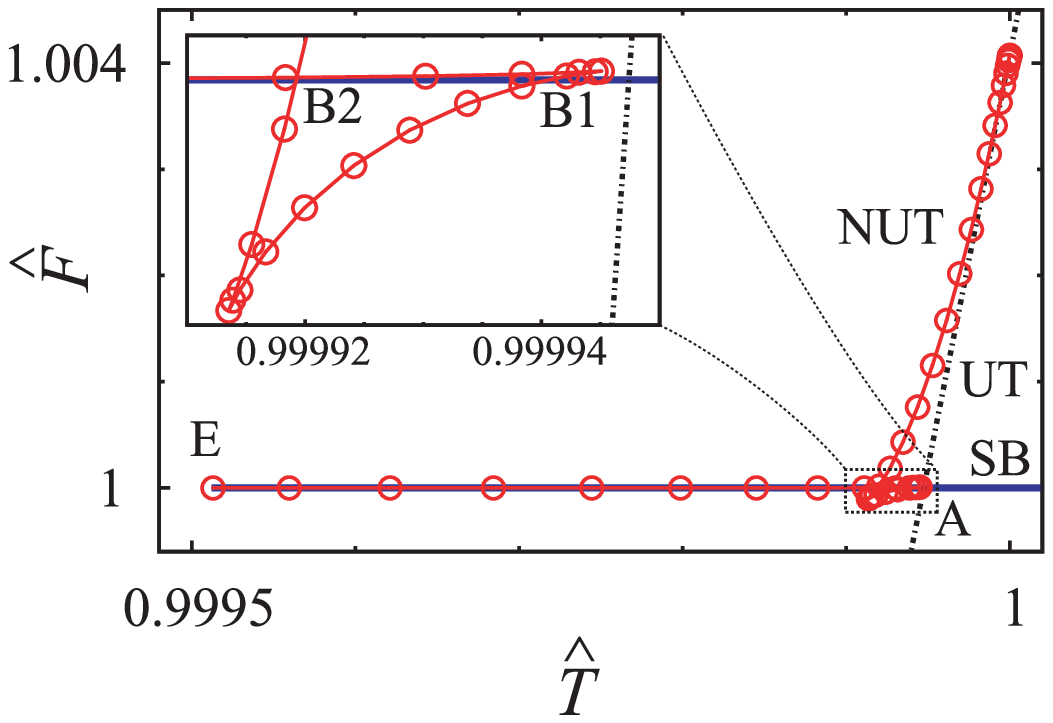} &
			\includegraphics[width=5.5cm]{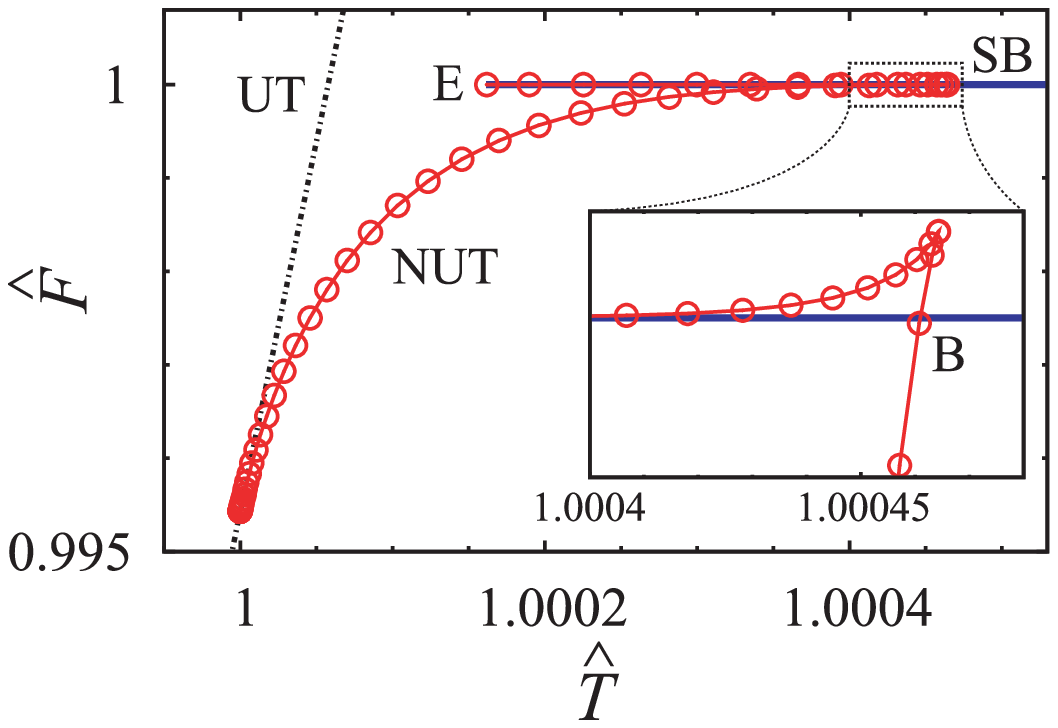} \\
		\end{tabular}
\caption{\textsf{The temperature-free energy diagrams for (a) $d=5$, (b) $d=10$, and (c) $d=11$, containing the phases of spherical ball (SB, blue-solid line), uniform tube (UT, black-dashed curve), and non-uniform tube (NUT, red-solid curve with small circles of data).
The phase having the smallest free energy is favored for a given temperature.
There is no cusp for $d=5$ (in fact, $4\leq d \leq 9$), while there exist
two cusps for $d=10$. One cusp is found in $d=11$ (in fact, $d \geq 11$).}}
\label{fg:(t,f)}
	\end{center}
\end{figure}

In the canonical ensemble, 
according to the number of cusps appearing in the ($\hat{T},\hat{F}$) diagram,
we again divide the dimensions into three groups
($4 \leq d \leq 9$, $d=10$, and $d \geq 11$).
The temperature-free energy diagrams for $d=5$, 10, and 11 are shown in Fig.~\ref{fg:(t,f)}.
Note that as mentioned before, the spherical ball phase 
has a lower bound of temperature, which is realized when the ball touches itself
 at the boundary. One can see how the transitions between the three phases 
occur by repeating discussions similar to those in the previous subsection.

In $d=5$, if we begin with a fat uniform tube (corresponding to a small 
$\hat{T}$) and increase $\hat{T}$, the uniform tube transits to 
a spherical ball at the point A, which is a first-order transition.
The non-uniform tube is not involved in this transition.
This transition pattern in the canonical 
ensemble is similar to that in the Kaluza-Klein black hole-black string 
system of low dimensions\footnote{We could not find literature
in which the temperature-free
energy diagram of the black hole-black string system is presented based on 
actual numerical data. But, see Fig.~1 in~\cite{AMMW} for a temperature-free 
energy diagram in 10 dimensions speculated from a perturbation analysis, 
which exhibits a similar behavior to Fig.~\ref{fg:(t,f)}(a).}.

The ($\hat{T},\hat{F}$) curve of the non-uniform tube in $d=10$
has two cusps. However, the non-uniform 
 tube branch is always above the uniform  tube branch. Therefore, the 
non-uniform tube branch has nothing to do with a realistic transition,
as the $d=10$ case in the microcanonical ensemble. That is, only the first-order transition from a uniform  tube to a spherical ball occurs (at the point A).

In $d=11$, if we begin with a fat uniform
tube and increase the temperature, the uniform tube meets the \RP critical 
point, located at $\hat{T}=1$, and at this point it transits smoothly 
to the non-uniform tube,
which has a smaller free energy than the uniform tube phase. This transition is of second or higher order. As the temperature increases further,
the non-uniform tube meets a spherical ball at the point B.
At this point, it transits to a spherical ball via a first-order transition.

In the canonical ensemble, the critical dimension at and above which
the smooth transition from the uniform to non-uniform tube phases is realized is 
$d_\ast^{\mathrm{can}}=11$. Hence, the critical dimension of the dual black 
strings is expected to be around $D_\ast^{\mathrm{can}}=d_\ast^{\mathrm{can}}+2
=13$.

The patterns of the phase transitions in both microcanonical and canonical ensembles
are summarized in Table~\ref{tbl:summary}.

\begin{table}[b!]
	\begin{center}
\caption{\textsf{{Diverse patterns of the phase transitions among the uniform tube
(UT), non-uniform tube (NUT), and spherical ball (SB) in the microcanonical
 and canonical ensembles. The `2nd' means that the order of transition may be either second or higher.}}}				
	\vskip 7pt
\footnotesize{
	\setlength{\tabcolsep}{15pt}
		\begin{tabular}[c]{cc|ccccc}
\hline \hline
\multicolumn{2}{c||}{Spacetime Dimension on Fluid Side: $d=n+3$}
&
\multicolumn{1}{c|}{4 - 9}
&
\multicolumn{1}{|c|}{10}
&
\multicolumn{1}{|c|}{11}
&
\multicolumn{1}{|c|}{12}
&
\multicolumn{1}{|c}{13 - $\ldots$}
\\
\hline
\hline
&
\multicolumn{1}{|c||}{Transition Type}
&
\multicolumn{2}{c}{UT $\longrightarrow$ SB}
&
\multicolumn{3}{|c}{UT $\longrightarrow$ NUT $\longrightarrow$ SB}
\\
\cline{2-7}
Microcanonical
&
\multicolumn{1}{|c||}{Order of Transition}
&
\multicolumn{2}{c}{1st}
&
\multicolumn{2}{|c}{1st \& 1st}
&
\multicolumn{1}{|c}{`2nd' \& 1st}
\\
\cline{2-7}
&
\multicolumn{1}{|c||}{No.~of Cusps in $(\hat{E},\hat{S})$}
&
\multicolumn{1}{c}{0}
&
\multicolumn{3}{|c}{2}
&
\multicolumn{1}{|c}{1} \\
\hline
\hline
&
\multicolumn{1}{|c||}{Transition Type}
&
\multicolumn{2}{c}{UT $\longrightarrow$ SB}
&
\multicolumn{3}{|c}{UT $\longrightarrow$ NUT $\longrightarrow$ SB}
\\
\cline{2-7}
Canonical
&
\multicolumn{1}{|c||}{Order of Transition}
&
\multicolumn{2}{c}{1st}
&
\multicolumn{3}{|c}{`2nd' \& 1st}
\\
\cline{2-7}
&
\multicolumn{1}{|c||}{No.~of Cusps in $(\hat{T},\hat{F})$}
&
\multicolumn{1}{c}{0}
&
\multicolumn{1}{|c}{2}
&
\multicolumn{3}{|c}{1} \\
\hline
\hline
\multicolumn{2}{c||}{Spacetime Dimension on Gravity Side: $D=d+2$}
&
\multicolumn{1}{c|}{6 - 11}
&
\multicolumn{1}{|c|}{12}
&
\multicolumn{1}{|c|}{13}
&
\multicolumn{1}{|c|}{14}
&
\multicolumn{1}{|c}{15 - $\ldots$}
\\
\hline
\hline
\end{tabular}
}
\label{tbl:summary}
\end{center}
\end{table}

\section{Summary and Discussion}
\label{sec:conclusion}

Adopting the equation of state for the fluid obtained by the 
Scherk-Schwarz compactification of ($d+1$)-dimensional conformal field
theory, we have investigated the thermodynamic properties of the `deconfined 
gluon plasma' lumps, which are expected to be dual to black holes and strings localized in the IR
of AdS${}_{d}$ on the Scherk-Schwarz and Kaluza-Klein circles.
We have invoked the fluid/gravity correspondence in order to predict the phase diagrams of the AdS black holes and strings.
We have found that those phase diagrams are qualitatively similar to 
those of the black hole-black string system in
the asymptotically locally flat Kaluza-Klein space. 
These results are not surprising in the sense that it had been known
that the phase structure of the black hole-black ring system in
AdS${}_{5}$ (AdS${}_{6}$) with Scherk-Schwarz compactification
is qualitatively similar to that of the black hole-black ring system
 in the 5(6)-dimensional flat background~\cite{plasmaring1,plasmaring2}. 
However, there is as yet no clear explanation 
for the agreement between the phase structures 
of such distinct systems. 
Furthermore, the critical dimensions found in this paper
($D_\ast^{\mathrm{microcan}}=15$ and $D_\ast^{\mathrm{can}}=13$) are `very 
close' (indeed equal in the canonical ensemble) to the ones in the 
asymptotically locally flat Kaluza-Klein space 
($D_{\ast\rm KK}^{\mathrm{microcan}}=
14$ and $D_{\ast\rm KK}^{\mathrm{can}}=13$) in the respective
 ensembles~\cite{Sorkin,KudohMiyamoto}. 
This closeness/coincidence, respectively, may
 stem from a universality of critical dimension, or 
there may exist other unknown reasons. 
As mentioned in the Introduction, the bulk duals may have a
non-trivial dependence on the holographic radial coordinate $u$. Furthermore,
the distribution of the size of the Scherk-Schwarz circle on horizon
cannot be determined \textit{a priori} before solving the Einstein equations. However, 
the $u$-dependence as well as the distribution of $\theta $-circles 
may be simply uniform in the limit of 
the \textit{large} black holes/strings for some reason, and in this case the 
phase structures would resemble those of the Kaluza-Klein system. 
It would be interesting to clarify 
the reason for the qualitative agreement, 
by solving the Einstein equations to find the gravity duals explicitly.

Next, we comment on the validity of our calculation, in particular, under what conditions the effective fluid description of the field theory is valid.
One condition is that the length scale over which the temperature and pressure 
vary must be
larger enough than the mean free path of quasiparticles of the system, 
which is of the same order as the inverse of the deconfinement temperature, 
$ T_c^{-1} $ (in the large 't Hooft coupling limit)
\cite{fluid-gravity1,plasmaring1}.
In our system, the temperature and
pressure are constant throughout the configurations, and this condition 
holds. Another condition
is that the temperature must not be far from the critical
temperature.
Otherwise, it is no longer valid to assume that the surface tension is a constant, whose value was implicitly assumed to be that at the critical temperature $\sigma=\sigma(T_c)$ in this paper.
From Fig.~\ref{fg:(t,f)}, we find that
the temperature of non-uniform tubes always satisfies
$T_\NUT/T_{\rp} = O(1)$. Since we have set $T_{\rp} \gtrsim T_c$ (by taking 
$\tilde{l}_0 \ll 1$), 
$ T_\NUT / T_c = O(1) $ holds for all configurations near the transition 
points we are interested in. 
Finally, the radius of curvature of the fluid surface in all directions must be much larger than the thickness of the surface, which is of order $ T_c^{-1} $.
If this does not hold, 
higher-derivative contributions to the surface stress tensor 
must be included.
Since we have discussed the case
$ L \gg l_0 = \sigma/\rho_0 \sim T_c^{-1}$,
we can safely ignore the thickness of the surface. However, the 
waist of non-uniform tube can become arbitrarily thin in the limit $\lambda=r_-/r_+ \to 0$,
where the thickness of the surface cannot be ignored. 
Thus, the transition process from a non-uniform tube to a
spherical ball could be modified by higher-derivative contributions.
One may wonder whether this problem can be resolved by 
a field theoretic approach.

One may object that the critical dimensions found in this paper are too 
large to occur in a realistic setting such as M/String theory,
and cannot describe reality.
However, a critical dimension at and above which a stable non-uniform phase appears 
will depend on various factors of the system under consideration. For instance, 
the critical
 dimension for boosted black strings is expected to be significantly lower 
for highly-boosted strings~\cite{Hovdebo}. Taking into account the fact that a
 large black ring~\cite{BR} is regarded as a boosted black string, it is
 expected that there is a \textit{stable undulating} black ring phase even in low 
dimensions such as 5 and 6. Therefore, it would be quite interesting to investigate
 the existence of the \RP instability for rotating plasma rings~\cite{plasmaring1,plasmaring2}
as well as its possible nonlinear consequences\footnote{See \cite{Evslin:2008py} for an interesting discussion on the (non-)existence of Saturn like configuration of plasma and its implications on the gravity side. We would like to thank one of the authors, C.~Krishnan, for informing us of his paper.}. 
In a similar vein, it would be also 
interesting to investigate the \RP instability of ultra-spinning plasma balls
(see~\cite{ultra-spinning} for a related observation on the gravity side), and 
to examine the (in)stability of the non-uniform tube phases obtained in this 
paper. These analyses may provide holographic interpretations of the 
stability of higher dimensional black holes.

\subsection*{Acknowledgments}

The authors would like to thank M.W.~Choptuik, O.J.C.~Dias, R.~Emparan, 
C.~Eling, P.~Figueras, G.W.~Gibbons, D.~Gorbonos, R.~Gregory, J.~Hansen,
T.~Harmark, M.~Koiso, B.~Kol, Y.~Mandelbaum, N.A.~Obers,
B.W.~Palmer, H.S.~Reall, E.~Sorkin, J.H.~Traschen, T.~Wiseman, and 
D.~Yamada for valuable conversations or comments on this and the related works.
 U.M.~thanks the organizers of the workshop ``Black Holes: A Landscape of 
Theoretical Physics Problems'' (25 August - 5 October 2008, CERN), where this 
work was finalized with the enjoyable discussions with the participants. 
They would also like to acknowledge
hospitality during their stay in September 2008 at DAMTP and the Centre for Theoretical 
Cosmology, Cambridge University.
K.M.~is supported in part by the Grant-in-Aid for Scientific Research Fund 
of the JSPS (No.~19540308) and for the Japan-U.K. Research Cooperative 
Program, and by the Waseda University Grants for the Special Research 
Projects. U.M.~is supported by the Golda Meir Fellowship, by 
the Israel Science Foundation Grant (No.~607/05), and by the DIP Grant 
(No.~H.52).

\appendix

\section{Manipulations for Accurate Numerics}
\label{sec:numerics}

The integrands in Eq.~(\ref{eq:NUB-quant}) diverge since $U(w_\pm)=0$. Therefore, one needs some manipulations to numerically integrate them accurately, which we describe here.

The  variables in Eq.~(\ref{eq:NUB-quant}), $\tilde{L}(\lambda)$, $\tilde{A}(\lambda)$, and $\tilde{V}(\lambda)$, can be written as
\be
	\tilde{X} (\lambda) = \int^{w_+}_{w_-} \dd w \frac{  \psi_X(w)  }{ \sqrt	{  -U(w)} },
\;\;\;
	(X=L,A,\mbox{or }V),
\label{eq:singular-integral}
\ee
where $\psi_X(w)$ is a certain regular function, i.e., $ \psi_L(w) := 2 $, $\psi_A(w) := 2 w^n \sqrt{ 1-U(w) }$, or $\psi_V(w) := 2 w^{n+1}$. Using the explicit form of the potential $U(w)$, one can write the integral as
\be
	\tilde{X} (\lambda)
	=
	\int^{w_+}_{w_-} \dd w
	\frac{  ( w^{n+1} + K ) \psi_X(w) }
	{ \sqrt{ ( w^n + w^{ n+1 } + K ) ( w_+ - w ) ( w - w_- ) g(w)  } },
\ee
where we have defined $(n-1)$-st order polynomial $g(w)$ by
\be
	g(w)
	=
	\sum_{ m=0 }^{ n-1 } g_m w^m
	:=
	\frac{ w^{n+1} - w^n + K }{ ( w - w_+ ) ( w - w_- ) }.
\label{eq:g-definition}
\ee
Comparing the coefficients of both sides of Eq.~(\ref{eq:g-definition}), we have a recursion relation of $g_m$,  
\be
&&
	g_m - ( w_+ + w_- ) g_{m+1} + w_+ w_- g_{m+2} = 0,
\;\;\;
	(0 \leq m \leq n-3),
\nonumber
\\
&&
	g_0 = \frac{ K }{ w_+ w_- },
\;\;\;
	g_1 = \frac{ ( w_+ + w_- ) K }{ (w_+ w_-)^2 },
\;\;\;
	g_{n-2} = w_+ + w_-  - 1,
\;\;\;
	g_{n-1} = 1.
\label{eq:recursion-BC}
\ee
This recursion relation is solved as
\be
	g_m
	=
	\frac{ g_{n-2} - w_- g_{n-1} }{ w_+ - w_- } w_+^{ n-1-m }
	-
	\frac{ g_{n-2} - w_+ g_{n-1} }{ w_+ - w_- } w_-^{ n-1-m },
\;\;\;
	( 2 \leq m \leq n-3 ).
\ee
Using the explicit expressions of $K$ and $w_\pm$, and Eq.~(\ref{eq:recursion-BC}), we obtain the final expression,
\be
	g_m
	=
	( 1-\lambda^{ m+1 } )
	( 1-\lambda^n )^{ n-2-m }
	\left(
		\frac{ \lambda }{ 1 - \lambda^{n+1} }
	\right)^{ n-1-m },
\;\;\;
	( 0 \leq m \leq n-1 ).
\label{eq:g-solution}
\ee
Then, we change the integration variable by
\be
	w
	=
	\frac{ w_+ + w_- }{ 2 }
	+
	\frac{ w_+ - w_- }{ 2 } \xi.
\label{eq:variable-change}
\ee
Thus, the integral~(\ref{eq:singular-integral}) is written as
\be
	\tilde{X} (\lambda)
	=
	\int_{-1}^{1} \dd \xi
	\frac{ ( w^{n+1} + K ) \psi_X(w) }{ \sqrt{ (w^n + w^{n+1} + K)  
 	( 1-\xi^2 ) g(w) } },
\label{eq:X-NUT}
\ee
where $w$ is regarded as the function of $\xi$ by 
Eq.~(\ref{eq:variable-change}), and $g(w)$ is given as the polynomial 
through Eqs.~(\ref{eq:g-definition}) and (\ref{eq:g-solution}). Finally, 
we use the following expressions for $(K,w_\pm)$ to reduce the round-off error, 
which are equivalent to the original definitions~(\ref{eq:Cw}),
\be
	w_+
	=
	\frac{ \sum_{m=0}^{n-1} \lambda^m }
	{ \sum_{m=0}^{n} \lambda^m },
\;\;\;
	w_-
	=
	\frac{ \sum_{m=0}^{n-1} \lambda^{m+1} }
	{ \sum_{m=0}^{n} \lambda^m },
\;\;\;
	K
	=
	\frac{ \left( \sum_{m=0}^{n-1} \lambda^{m+1} \right)^n }
	{ \left( \sum_{m=0}^{n} \lambda^m \right)^{n+1} }.
\label{eq:Cw2}
\ee
We find that the expressions (\ref{eq:X-NUT}) and (\ref{eq:Cw2}) significantly reduce the numerical errors and make it possible to obtain non-uniform tube solutions, even for $\lambda \ll 1$.
A {\sc mathematica} notebook is available at the author's website,
{\tt http://www.phys.huji.ac.il/$\verb+~+$umpei/}.

\section{Justification and Interpretation of Cusp Structures}
\label{sec:justification}

\begin{table}[b!]
	\begin{center}
\caption{\footnotesize{\textsf{Numerical values of the slenderness parameter, Eq.~(\ref{eq:slender}).
}}}
\label{table:values}
\vspace{6pt}
\setlength{\tabcolsep}{8pt}
\footnotesize{
\begin{tabular}{  c  c  c  c  c  c  c  c  c  c  c  c }
 \hline \hline
 $d=n+3$
 & 4 & 5 & 6 & 7 & 8 & 9 & 10 & 11 & 12 & 13 & 14
 \\
 \hline
 $ \Lambda $
 &	$3.142$		&	$2.221$		&	$1.814$
 &	$1.571$		&	$1.405$		&	$1.283$		&	$1.187$		&	$1.111$
 &	$1.047$		&	$0.9935$ 	&	$0.9472$			
 \\  
 \hline\hline
	\end{tabular}
}
	\end{center}
\end{table}

Here, we argue that the cusp structures observed in the phase diagrams are not numerical artifacts. The phase curves were obtained by numerical integration of Eq.~(\ref{eq:X-NUT}), which was done with $\mathtt{NIntegrate}$ in the {\sc mathematica} code. We confirmed the convergence of integrated values by changing the parameters of $\mathtt{AccuracyGoal}$, $\mathtt{PrecisionGoal}$, and $\mathtt{WorkingPrecision}$. That is, the digits required to draw the fine structures of phase diagrams such as the cusp structure do not change against the change of these parameters. It is noted that such a convergence is made rapid by virtue of the manipulations described in Appendix~\ref{sec:numerics}. The fact that the curve of non-uniform tube phase approaches the \RP critical point and the tip of the spherical ball phase at the ends of curve supports our numerics independently.

We also note that the existence of cusp structures, especially during the transient dimensions discussed in Secs.~\ref{sec:class2} and \ref{sec:cano}, is not inconsistent with our conclusion that the phase structure of black strings and black holes in the Scherk-Schwarz compactified AdS is similar to that of the black hole-black string system in the asymptotically locally flat Kaluza-Klein space. That is, the critical dimensions known in the Kaluza-Klein black hole-black string system~\cite{Sorkin,KudohMiyamoto} were predicted by higher-order perturbations, which tells us the behaviors of a phase curve only near the \GL critical point. Therefore, although the critical dimension of Kaluza-Klein black string has been known to be $D_{\ast,\mathrm{KK}}=14$ in the microcanonical for example, the possibility is not excluded that the non-uniform black string branches in $D_{\mathrm{KK}}=12$ and $D_{\mathrm{KK}}=13$ behave like the curves of non-uniform tube in Fig.~\ref{fg:10d(e,s)} (note that the fully non-linear behaviors of non-uniform black string phase curves have been obtained for $D_{\mathrm{KK}} \leq 11$ in~\cite{Sorkin2}, but such a behavior was not observed). Therefore, the cusp structures in the transient dimensions found in this paper are not only consistent with the known gravitational calculations but also suggest an important lesson that a critical dimension should be determined \textit{non-perturbatively}, or lesson that the critical dimension defined by the smooth transition from the uniform branch to non-uniform branch is not enough to know the global structure of a phase diagram. It would be interesting to perform numerical analyses to obtain the fully non-linear behaviors of non-uniform black strings for $D_{\mathrm{KK}}=12$ and $13$, although their difficulty was one of the original motivations to work on fluids in this paper.

In Ref.~\cite{MiyamotoMaeda}, we found the critical dimension at and above which the non-uniform tube branch emanates from the \RP critical point with decreasing volume $V$ for a fixed period $L$, of which existence is related to the critical dimension in the thermodynamic phase diagrams obtained in this paper. We found a simple criterion whether the non-uniform branch increases or decrease its volume near the \RP critical point as follows.
Let us define a `slenderness' parameter of the critical uniform tube,
\be
	\Lambda
	:=
	\frac{ L_\rp }{ 2r_0 }
	=
	\frac{\pi}{\sqrt{n}}.
\label{eq:slender}
\ee
We may say that a critical tube is slender, say, if $\Lambda>1$, while one is fat if $\Lambda<1$.
The numerical values of this parameter are given in Table.~\ref{table:values}.
From the table, the critical tube is found to be slender for $4 \leq d\leq 12$, while fat for $d \geq 13$.
On the other hand, the non-uniform tube branch increases its volume for $4 \leq d \leq 12$, while decreases it for $d \geq 13$ near the critical point (see Figs.~2 and 4 in~\cite{MiyamotoMaeda}).
The above agreement of two threshold dimensions seems not to be just a coincidence from a simple geometric consideration:
the non-uniform tube branch emanating from a slender critical tube has to increase its volume for a fixed period $L$ in order to approach a spherical ball, while one emanating from a fat critical tube has to decrease its volume.
Thus, the slenderness provides us a simple criterion to forecast the behavior of non-uniform tube around the \RP critical point, and our numerical results support this geometric expectation and fit our intuition.

Finally, it should be stressed that the transient behaviors of non-uniform tube observed in Fig.~\ref{fg:10d(e,s)} (and also Fig.~\ref{fg:(t,f)}) are quite natural in that the drastic change of phase structure between the lower dimensional class ($4 \leq d \leq 9$) and higher dimensional class ($d\geq 13$) was `smoothly interpolated' by the appearance of the `swallowtail' (i.e., the part of the curve between the pair of two cusps) at $d=10$ and its subsequent growth at $d=11,12$. It would be interesting to regard dimension $d$ as a fictitious continuous parameter and perform the same calculations to obtain phase diagrams for non-integer `dimensions' $ 9 \leq d \leq 13 $, which would show us a continuous change of phase structure.



\end{document}